\documentclass[preprint,12pt,authoryear]{elsarticle}

\newif\ifarxiv
\arxivtrue 
\ifarxiv
  \makeatletter
  \def\ps@pprintTitle{%
    \let\@oddhead\@empty
    \let\@evenhead\@empty
    \let\@oddfoot\@empty
    \let\@evenfoot\@oddfoot}
  \makeatother
\else
  \journal{<Elsevier Journal Name>} %
\fi

\usepackage{amssymb}
\usepackage{amsmath}
\usepackage{url}
\usepackage{multirow}
\usepackage{subcaption}
\usepackage{float}
\usepackage[section]{placeins}
\usepackage{booktabs}
\usepackage{listings}
\usepackage{xcolor} %
\usepackage[normalem]{ulem}
\usepackage{hyperref}
\hypersetup{colorlinks=true} %

\usepackage{natbib}
\usepackage{siunitx}

\journal{Neurocomputing}

\begin{document}

\begin{frontmatter}

\title{Time Series Analysis of Spiking Neural Systems via Transfer Entropy and Directed Persistent Homology}

\author[uon]{Dylan Peek} %
\author[cmi]{Siddharth Pritam} %
\author[rmit]{Matthew P. Skerritt} %
\author[uon]{Stephan Chalup} %

\affiliation[uon]{organization={The University of Newcastle},
            addressline={School of Information and Physical Sciences}, 
            city={Newcastle},
            postcode={2308}, 
            state={New South Wales},
            country={Australia}}

\affiliation[cmi]{organization={Chennai Mathematical Institute},
            addressline={Computer Science Group}, 
            city={Chennai},
            postcode={603103}, 
            state={Tamil Nadu},
            country={India}}

\affiliation[rmit]{organization={The Royal Melbourne Institute of Technology},
            addressline={Dept. of Mathematical and Geospatial Sciences}, 
            city={Melbourne},
            postcode={3000}, 
            state={Victoria},
            country={Australia}}

\begin{abstract}
We present a topological framework for analysing neural time series that integrates Transfer Entropy (TE) with directed Persistent Homology (PH) to characterize information flow in spiking neural systems. TE quantifies directional influence between neurons, producing weighted, directed graphs that reflect dynamic interactions. These graphs are then analyzed using PH, enabling assessment of topological complexity across multiple structural scales and dimensions.

We apply this TE+PH pipeline to synthetic spiking networks trained on logic gate tasks, image-classification networks exposed to structured and perturbed inputs, and mouse cortical recordings annotated with behavioral events. Across all settings, the resulting topological signatures reveal distinctions in task complexity, stimulus structure, and behavioral regime. Higher-dimensional features become more prominent in complex or noisy conditions, reflecting interaction patterns that extend beyond pairwise connectivity.
Our findings offer a principled approach to mapping directed information flow onto global organizational patterns in both artificial and biological neural systems. The framework is generalizable and interpretable, making it well suited for neural systems with time-resolved and binary spiking data.
\end{abstract}


\end{frontmatter}

\section{Introduction}
Understanding how information flows through neural systems remains a central challenge in neuroscience and artificial intelligence. In this work, we adopt a topological perspective to explore how functional neural activity changes under different task conditions. Our approach is motivated by the idea that population-level neural responses may organize into higher-order structures that reflect more than just pairwise interactions. 

To investigate topological structures of neural time-series data, we utilize a framework that combines Transfer Entropy (TE)~\citep{schreiber2000measuring} and Persistent Homology (PH)~\citep{EdelsbrunnerHarer2010, carlsson2009topology}. TE is an information-theoretic measure that captures the causal influence from one neuron to another, resulting in a weighted and directed network of interactions. Unlike structural connectivity, TE reflects dynamic, functional relationships that evolve with task demands.

We then apply PH to these TE-derived graphs. PH is a method from Topological Data Analysis (TDA)~\citep{otter2017roadmap} that tracks how features such as connected components, loops, and higher-dimensional cavities emerge and disappear across a filtration, such as a weighted-edge threshold.

We evaluate the framework in three settings: (1) synthetic spiking networks that implement logic-gate computations, (2) image-classification spiking networks under structured and perturbed pixel inputs, and (3) mouse cortical recordings aligned to annotated behavioural events~\citep{ibl2021neuro}. Across these domains, we observe:
\begin{itemize}
\item consistent, distinguishable topological distributions of directed information flow for each task, even with fixed architecture and weights;
\item richer higher-dimensional homology for more complex or noisier tasks, such as XOR relative to AND/OR and high-noise digits relative to clean inputs;
\item Wasserstein distances between persistent-feature distributions that recover meaningful geometry in task space, including ordered transitions across noise levels.
\end{itemize}

Methodologically, the pipeline is generalisable, interpretable, and computationally tractable for binary spike trains and time-resolved neural data, which enables application to both artificial and biological systems. Conceptually, it maps directed information flow to emergent topological organisation and provides a lens to characterise task dynamics and compare network regimes.

Our central claim is that task-driven spiking activity exhibits emergent topological and geometric structure which correlates with stimulus structure and task difficulty. 

This work offers a cross-disciplinary foundation for applying topological data analysis to neural computation. It includes the necessary background, 
and contributes a reproducible methodology for linking directed information flow to topological organization in both artificial and biological systems.

\section{Literature Review}
The topological analysis of time-series signals is an emerging technique with a focus on uncovering the underlying structure of dynamic systems. In this review, we first outline applications of TDA in neuroscience, followed by artificial settings such as recurrent and spiking neural networks. We then examine existing approaches and key literature that connect time series, functional connectivity, and directed or information-theoretic constructions. Finally, we position our study within this landscape by outlining how our contributions build on and extend these threads.

\subsection{Topological Data Analysis in Neuroscience}

Topological Data Analysis (TDA) has gained traction as a tool to characterize complex neural systems by extracting shape-based features from high-dimensional data. One of the earliest applications came from \citet{dabaghian2012topological}, who used persistent homology (PH) to analyze hippocampal place cell activity, demonstrating that neural co-firing patterns encode the topological structure of spatial environments. This work introduced the concept of cognitive maps as topological rather than metric constructs.

\citet{petri2014homological} extended this idea to functional magnetic resonance imaging (fMRI), applying PH to time-varying correlation networks of the human brain. Their concept of ``homological scaffolds'' revealed higher-order structures such as loops and voids, undetectable by conventional graph analysis. Similar insights were made by \citet{giusti2015clique}, who used clique complexes to uncover geometric structure in neural correlations, suggesting that collective firing patterns may lie on low-dimensional manifolds with non-trivial topology.

Subsequent work by \citet{kang2021cohomology} used persistent cohomology on simulations of the hippocampal spatial representation system to evaluate conditions for successful topological discovery. They showed when neural firing data can reliably recover the known spatial topology, thereby offering practical guidelines for applying TDA in neural recordings.

\citet{reimann2017cliques} analyzed cortical microcircuits using algebraic topology, identifying high-dimensional cliques and cavities formed by structured synaptic connections. Their work highlighted the potential of directed topological features to track stimulus processing and neural coordination in space and time. 

In parallel, \citet{saggar2018dynamical} introduced a TDA-based approach to whole-brain fMRI dynamics, using the Mapper algorithm to capture an interactive representation of brain states without collapsing data across time. This revealed fine-grained within- and between-task transitions.

\subsection{Topological Characterization of Artificial Neural Dynamics}

Beyond biological recordings, TDA has been used to characterize artificial recurrent and spiking neural networks. \citet{guidolin2022geometry} demonstrated that cortical spike trains exhibit classifiable topological features under visual stimulation, providing a bridge between real-world sensory input and topological representation.

In deep learning, \citet{naitzat2020topology} examined how neural networks transform data topology layer-by-layer, showing that persistent homology can describe representational change through training. These approaches support the use of topology not just to describe structure, but to interpret dynamic computation. 

In simulated spiking circuits, \citet{bardin2019topological} demonstrated that persistent homology features extracted from spike train metrics can classify different network activity regimes. By building simplicial complexes on spike-train distance matrices, they distinguished oscillatory patterns and even generalized to unseen networks, highlighting the discriminatory power of topological features.

Similarly, \citet{bai2024transitions} computed Betti curves from the spike activity of a stochastic neuron network and used them to accurately discriminate sub-critical, critical, and super-critical states, suggesting that topological complexity changes at phase transitions. This study underscore how TDA can reveal latent dynamical structure or phase changes in synthetic neural systems, beyond what standard metrics capture.

\citet{muller2023topological} examined reinforcement learning agents with feed-forward networks by constructing sparse functional graphs from continuous single-pass activations using an activation and similarity threshold. Applying persistent homology to these graphs, they tracked the emergence of higher-order Betti numbers during training.

\subsection{TDA of Time Series and Functional Connectivity}

TDA has also been applied to analyze time-dependent networks derived from multivariate time series. \citet{perea2015sliding} introduced a sliding window embedding technique that converts univariate signals into geometric loops in high-dimensional space, quantifiable via persistent homology. ~\citet{stolz2017persistent} extended this to temporal networks, computing Betti curves over evolving graph filtrations.

\citet{myers2019persistent} and \citet{santos2019topological} used PH to detect state transitions in systems like EEG, showing that topological phase transitions reflect underlying dynamical changes. These studies demonstrate how topological features such as loops and voids serve as indicators of system-level reconfiguration.

\subsection{Information-Theoretic and Directed Graph Approaches}

Most TDA pipelines to date apply persistent homology to undirected graphs derived from correlation or similarity matrices. However, several works have explored directed constructions. \citet{chowdhury2018persistent} proposed a Dowker-based persistent homology pipeline applied to correlated place-cell activity. They retained directionality and successfully recovered environmental topology from inferred information flow. 

\citet{reimann2017cliques} and \citet{sizemore2019importance} have also emphasized the importance of directionality and high-order interactions, introducing directed flag complexes and highlighting topological motifs that represent structured communication beyond pairwise links.

A recent study by \citet{XI2025130086} used directed persistent homology combined with transfer entropy (TE+PH) on EEG data to compare pre- and post-stroke patients. The analysis operates at the regional EEG scale rather than individual neurons or putative cell co-firing. This study identified a clear topological distinction between the groups.

\subsection{Manifold Hypothesis and Neural Dynamics}The manifold hypothesis posits that high-dimensional population activity often evolves within structured, lower-dimensional subspaces of neural state space \citep{cunningham2014dimensionality}. Empirically, non-trivial topology has been observed in such population activity—for example, toroidal structure in grid-cell ensembles \citep{gardner2022toroidal} and a circular topology encoding stimulus orientation in mouse visual cortex \citep{beshkov2024topological}. While manifold-centric studies often emphasize latent geometry, our approach focuses on \emph{directed information flow}: we estimate time-resolved, weighted directed graphs via transfer entropy and characterize their multi-scale organization using persistent homology on directed flag complexes.

\subsection{Comparison to Present Work}
Prior studies often treated spiking networks, topological encoding, and information flow separately. Few have combined all three; Only one recent work applied directed PH with information-theoretic coupling at an EEG scale~\citep{XI2025130086}. Here, we extend these threads by applying directed persistent homology to neuron-level transfer entropy networks. To our knowledge, no study has applied directed PH to fine-grained spiking networks using an information-theoretic (TE) construction.

We evaluate the approach in both synthetic spiking networks and biological recordings.
In the synthetic setting, we control inputs, outputs, and trainable tasks. This removes confounds and lets us test the method along with the link between topological complexity, dimensionality, and task demands. We also design experiments that separate task space into quantitative levels (noise) and categorical classes, with controlled network stimulus, to isolate how each factor shapes the resulting topology. We have full observability over the synthetic models to assess edge weight and spike train volume directly. However, unlike existing studies that utilize synthetic models, we limit our analysis to spike trains only via transfer entropy to ensure generalization across unobservable simulated or real-world applications. 

We also aim to offer a central reference that explains the TDA pipeline and the required mathematical and conceptual background to connect neuroscience, computational topology, and computer science.

These elements produce a reproducible, directed, information-theoretic TDA pipeline that connects the interpretability of synthetic SNNs with the richness of biological data. In practice, time-resolved multidimensional metrics in both simulated and real neural activity provide an interpretable toolkit for analyzing dynamics and comparing tasks across systems and scales.

\section{Background}
\subsection*{Artificial Spiking Neural Networks}

Artificial Spiking Neural Networks (SNNs) are biologically inspired models that encode and process information via discrete spike events rather than continuous activations~\citep{maass1997networks,ghosh2009spiking}. In contrast to conventional artificial neural networks (ANNs), SNNs incorporate a notion of time, with neurons evolving their internal states over multiple simulation steps and emitting spikes conditionally based on membrane potential dynamics. While the mechanisms of such networks don't attempt to simulate the complexities of biological neural systems, for our experiments they provide a synthetic, trainable architecture for studying information-flow dynamics. The ability to extract neuronal spike-train data allows the proposed TDA pipeline to generalize between synthetic and real-world spiking networks.   

In this study, we utilize a fully connected recurrent architecture based on the \textit{Leaky Integrate-and-Fire (LIF)} neuron model~\citep{gerstner2002spiking}. Each neuron $i$ maintains a membrane potential $v_i(t)$ that integrates incoming current over time and decays exponentially:

\begin{equation}
    v_i(t) = \beta \cdot v_i(t{-}1) + I_i(t)
\end{equation}

\noindent where $\beta \in [0, 1)$ is the membrane decay factor, and $I_i(t)$ is the total synaptic input current received at time $t$, typically computed as:

\begin{equation}
    I_i(t) = \sum_j w_{ji} \cdot s_j(t)
\end{equation}

\noindent Here, $w_{ji}$ denotes the synaptic weight from presynaptic neuron $j$ to postsynaptic neuron $i$, and $s_j(t) \in \{0, 1\}$ is the spike emitted by neuron $j$ at time~$t$.

When the membrane potential exceeds a fixed threshold (typically normalized to $1$), the neuron emits a spike:

\begin{equation}
    s_i(t) = H(v_i(t) - \theta), \quad \theta = 1
\end{equation}

\noindent where $H(\cdot)$ is the Heaviside step function. After spiking, the membrane potential is reset or reduced by subtracting the threshold, preserving sub-threshold potential dynamics.

\subsubsection*{Network Architecture and Input Encoding}

The networks used in this work are shallow, fully connected SNNs comprising a recurrent spiking block followed by weighted readout neurons. Inputs are encoded as spike trains via \textit{Poisson encoding}~\citep{gerstner2002spiking}, in which each input feature (e.g., binary bit or task channel) is mapped to a spike train over $T$ simulation steps:

\begin{equation}
    s^{(\text{in})}_{i}(t) \sim \text{Bernoulli}(p_i), \quad p_i \in \{p_{\text{high}}, p_{\text{low}}\}
\end{equation}

\noindent This frequency-based encoding scheme allows binary and categorical inputs to be represented with distinct spiking activity levels. Input features are concatenated to form the full input layer which connects to internal hidden neurons by fully-connected, trainable edges.

Output signals are decoded either from a fixed-length window of neuron activity (e.g., by applying a readout weight to the mean firing rate over that window) or via a dedicated output neuron with its own membrane dynamics accumulating spikes. In both cases, weighted connections from hidden neurons to the output layer determine the network’s response.

\subsubsection*{Training and Surrogate Gradient Descent}

Since spiking is non-differentiable, training is performed using \textit{surrogate gradient descent}~\citep{neftci2019surrogate}, which replaces the gradient of the Heaviside spike function with a smooth approximation during backpropagation. A common surrogate is the fast sigmoid:

\begin{equation}
    \frac{dH(v)}{dv} \approx \frac{1}{(1 + \alpha |v|)^2}
\end{equation}

\noindent where $\alpha$ controls the sharpness of the approximation. This allows the model to be trained end-to-end using variants of stochastic gradient descent such as Adam or AdamW.

To account for the temporal dynamics, the full sequence of membrane potentials and spikes is unrolled over $T$ time steps, enabling backpropagation through time (BPTT)~\citep{neftci2019surrogate}. While BPTT is computationally expensive, it remains the standard approach for training temporal SNNs, and efficient implementations leverage GPU acceleration and sparse updates.

\subsubsection*{Remarks on Expressivity and Temporal Computation}

The networks used in this study are capable of solving structured binary tasks (e.g., logic gates) and classifying temporally encoded visual data. The internal dynamics allow neurons to act as memory elements, capturing temporal correlations across the input spike trains. Output classification is performed via thresholding on output neuron activity (e.g., rate or final membrane value), enabling binary or continuous predictions.

\subsection{Transfer Entropy}
Transfer Entropy (TE) is an information-theoretic measure introduced by ~\citet{schreiber2000measuring} that quantifies the directed influence of one time series on another. Given two discrete time-series signals, $X_t$ and $Y_t$, TE from $X$ to $Y$ is defined as:

\begin{equation}
    T_{X \rightarrow Y} = \sum p(y_{t+1}, y_t^{(k)}, x_t^{(l)}) \log \frac{p(y_{t+1} \mid y_t^{(k)}, x_t^{(l)})}{p(y_{t+1} \mid y_t^{(k)})}
    \label{eq:te}
\end{equation}

\noindent where:
\begin{itemize}
    \item $X_t$ and $Y_t$ are discrete-time stochastic processes representing the source and target signals respectively.
    \item $y_{t+1}$ is the future state of $Y$ at time $t+1$.
    \item $y_t^{(k)} = \{y_t, y_{t-1}, \ldots, y_{t-k+1}\}$ is the $k$-length history (embedding) of the target process.
    \item $x_t^{(l)} = \{x_t, x_{t-1}, \ldots, x_{t-l+1}\}$ is the $l$-length history of the source process.
    \item $p(\cdot)$ denotes the joint or conditional probability mass function over the relevant variables.
\end{itemize}

\noindent Intuitively, $T_{X \rightarrow Y}$ quantifies the additional predictive power gained about $Y$'s next state by including the past of $X$, beyond what is already explained by $Y$'s own past. A non-zero TE implies a directed interaction or information transfer from $X$ to $Y$.

The computation of Transfer Entropy as given in Equation~\ref{eq:te} depends on the estimation of several joint and conditional probability mass functions (PMFs), such as $p(y_{t+1}, y_t^{(k)}, x_t^{(l)})$ and its marginals.

In discrete systems like binned spike trains, these PMFs are typically estimated via frequency-based counting methods over a sliding window. Let $\mathcal{D}$ denote the dataset of observed time-series values over $T$ time steps. The joint PMF can be estimated as:

\begin{equation}
    \hat{p}(y_{t+1}, y_t^{(k)}, x_t^{(l)}) = \frac{1}{T - \max(k,l)} \sum_{t = \max(k,l)}^{T-1} \mathbb{I}\left[(y_{t+1}, y_t^{(k)}, x_t^{(l)}) = (\tilde{y}, \tilde{y}^{(k)}, \tilde{x}^{(l)})\right]
    \label{eq:pmf}
\end{equation}

\noindent where:
\begin{itemize}
    \item $\mathbb{I}[\cdot]$ is the indicator function, equal to 1 when the condition holds, and 0 otherwise.
    \item $(\tilde{y}, \tilde{y}^{(k)}, \tilde{x}^{(l)})$ is a specific tuple of observed values for which the PMF is estimated.
    \item The denominator accounts for the total number of valid embedding windows.
\end{itemize}

\noindent From the joint PMFs, conditional probabilities can be computed via:

\begin{equation}
    \hat{p}(y_{t+1} \mid y_t^{(k)}, x_t^{(l)}) = \frac{\hat{p}(y_{t+1}, y_t^{(k)}, x_t^{(l)})}{\hat{p}(y_t^{(k)}, x_t^{(l)})}
\end{equation}

\noindent Estimation accuracy depends critically on the choice of:
\begin{itemize}
    \item \textbf{Embedding lengths} $k$ and $l$: Higher values capture more temporal structure but require more data.
    \item \textbf{Binning resolution}: The choice of time bin size can affect sensitivity to synchronous vs. asynchronous spiking.
    \item \textbf{Smoothing or regularization}: For sparse spike trains or small datasets, smoothing (e.g., adding pseudocounts) helps avoid zero-probability issues.
\end{itemize}

For binary spike trains, the state space is small, making frequency-based estimation feasible.

\subsection{Topology and Persistent Homology}

Topology is the mathematical study of shape and structure~\citep{EdelsbrunnerHarer2010}, focusing on properties that are preserved under continuous deformation such as stretching or bending, but not tearing or gluing. Unlike geometry, which depends on precise notions of distance and angle, topology concerns itself with global structure—how parts of a space are connected. For example, a sphere and a cube are topologically equivalent, as they can be deformed into each other without cutting or puncturing. These structures can be formally described by topological invariants such as connected components, holes, and voids.

To analyze such properties in discrete data, especially networks or point clouds, topologists construct \textit{simplicial complexes}: combinatorial objects built from vertices, edges, triangles, and higher-dimensional simplices. A $k$-simplex represents a $(k+1)$-clique in a graph—e.g., a 2-simplex is a triangle, a 3-simplex a tetrahedron. The set of all simplices satisfying specific connectivity rules defines a \textit{complex}.

In our setting, we apply these ideas to directed graphs constructed from neural data. Specifically, we use the \textit{directed flag complex}, an adaptation of the clique complex for asymmetric networks~\citep{lutgehetmann2020computing}. A directed $k$-simplex is formed when all directed edges (excluding loops and reciprocal edges) exist between a set of $k+1$ nodes in a consistent orientation. These complexes allow us to represent not only the presence of connections but also their directionality, capturing richer structure than undirected methods.

Persistent Homology (PH) is a tool from Topological Data Analysis (TDA) that summarizes how topological features—such as connected components (dimension 0), loops (dimension 1), and higher-dimensional cavities—appear and disappear across multiple scales~\citep{carlsson2009topology}. This is achieved by constructing a \textit{filtration}: a nested sequence of simplicial complexes generated by sweeping through a scalar function (e.g., edge weights) in order of increasing threshold. In our case, the filtration is based on \textit{inverted transfer entropy} values, so that strong information transfers appear earlier in the complex.

As the filtration progresses, topological features are born (appear) and eventually die (merge or vanish). Persistent homology tracks these events across dimensions and filtration values, recording each feature as a tuple $(b, d, d')$ denoting its birth time $b$, death time $d$, and homology dimension $d'$. The resulting multiset of such tuples forms the \textit{persistence diagram}, a geometric summary of the evolving topology.

A related concept is the \textit{Betti number}, $\beta_k$, which counts the number of $k$-dimensional topological features (e.g., $\beta_0$ for connected components, $\beta_1$ for loops) present at a given filtration threshold. Tracking these counts as a function of the threshold gives rise to the \textit{Betti curve} $\beta_k(\epsilon)$~\citep{nardini2021topological}. Integrating this curve over the filtration range yields a single scalar summary known as the \textit{area under the Betti curve} (AUBC), which we use in our analysis to quantify the total topological activity per dimension. These topological features, especially in higher dimensions, may reflect the presence of nontrivial cycles and cavities in the functional organization of neural activity consistent with the hypothesis that such activity evolves along manifold-like structures~\citep{reimann2017cliques}. 

By combining directed flag complexes with persistence diagrams and Betti-based summaries, we are able to characterize the evolving, multi-dimensional topology of information flow in neural systems in a principled and scalable manner.

\section{Topological Analysis Pipeline} \label{sec:pipeline}
\begin{figure}[ht]
  \centering
  \makebox[\linewidth][c]{%
    \includegraphics[width=1.2\linewidth]{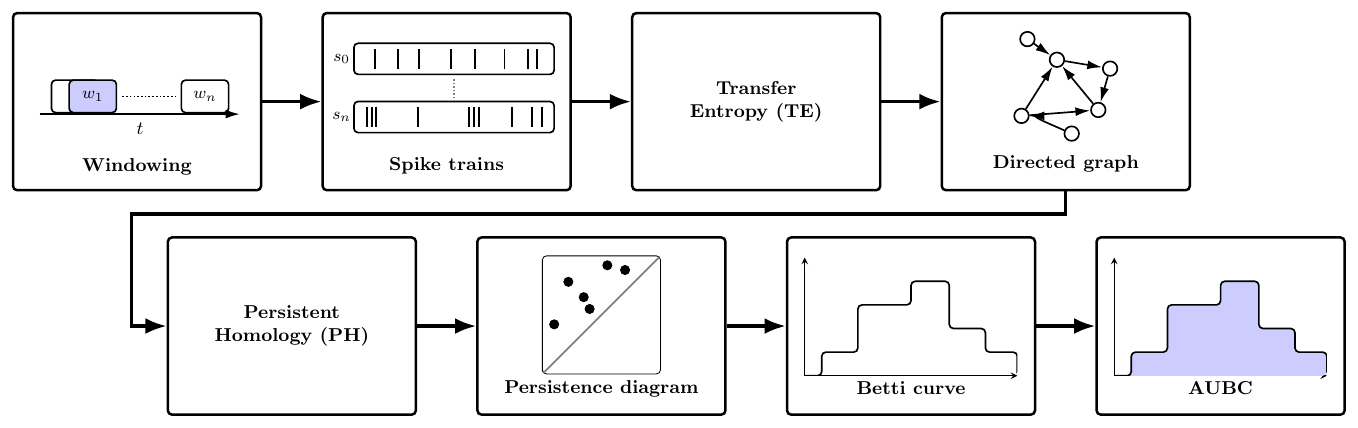}%
  }
  \caption{The complete topological analysis pipeline outlined in Section~\ref{sec:pipeline}.}
  \label{fig:pipeline}
\end{figure}

To characterize the topology of the information dynamics of spiking neural systems, we construct a pipeline that begins with raw spike trains and produces a compact, interpretable summary of the network's evolving information structure. This pipeline is applied consistently across all experiments, with only minor modifications to accommodate differences in network size, labeling regimes, or temporal granularity. A visual summary of the pipeline can be seen in Figure~\ref{fig:pipeline}.

\subsection*{Spike Train Acquisition and Preprocessing}

The input to the pipeline consists of discrete-time binary spike trains, each representing the activity of an individual neuron over time. These spike trains may originate from artificial spiking neural networks (SNNs) or biological recordings and are aligned to the experimental timebase, allowing temporal correspondence with stimuli or behavioral events.

In the case of long or continuous recordings, we segment the spike data into overlapping temporal windows. Each window spans a fixed duration (e.g., 50--250 ms) with a specified step size, allowing us to track temporal evolution of network topology. The choice of window size controls the temporal \textit{scope} of the analysis, while the step size determines its \textit{resolution}.

\subsection*{Transfer Entropy Estimation}

For each window (or entire dataset in the case of brief, discrete inputs), we compute the pairwise \textit{Transfer Entropy} (TE) between all neuron pairs. We use the \texttt{pyinform} library~\citep{moore2018inform}, which provides efficient and bin-based estimation of TE for discrete-valued time series.

TE quantifies the directed influence of one neuron's spiking history on the future activity of another, capturing causal interactions beyond what is explained by the target neuron's own past. The result is a weighted, directed adjacency matrix $A \in \mathbb{R}^{N \times N}$, where $A_{ij}$ denotes the TE from neuron $i$ to neuron $j$.

To improve computability and robustness, we apply two preprocessing steps to the TE adjacency matrix:
\begin{itemize}
    \item \textbf{Self-loop removal}: Diagonal elements $A_{ii}$ are set to zero. In addition to being structurally trivial, self-connections are not defined under transfer entropy, as a signal contains no information transfer to itself.
    \item \textbf{Asymmetry enforcement}: For each neuron pair $(i, j)$, we retain only the stronger directed connection. Specifically, we set $A_{ij} = \max(A_{ij}, A_{ji})$ and $A_{ji} = 0$ if $A_{ij} > A_{ji}$, or vice versa. This removes reciprocal links and ensures a unidirectional graph structure.
    \item \textbf{Inversion for sublevel filtration}: Because we use a \textit{sublevel} filtration in persistent homology, edge weights must be interpreted such that lower values correspond to stronger connections. We therefore invert the TE values via a monotonic transformation (e.g., $A_{ij} \rightarrow -A_{ij}$) so that edges with higher information transfer appear earlier in the filtration.
\end{itemize}

\subsection*{Directed Persistent Homology via Flagser}

We analyze the resulting TE graph using \textit{persistent homology} computed on directed flag complexes. Specifically, we employ the \texttt{Flagser} library \citep{lutgehetmann2020computing}, which supports persistent homology computations on directed, weighted graphs and permits arbitrary homology dimensions. 

To construct the filtration, we interpret the (preprocessed and inverted) TE adjacency matrix as a weighted directed graph, where lower edge weights represent stronger information transfer. A \textit{sublevel filtration} is then applied: edges are added in order of increasing weight, starting from those with the greatest transfer entropy. This produces a nested sequence of directed graphs, capturing increasingly weaker interactions.

At each stage of the filtration, directed simplices are included according to the rules of the directed flag complex: a $k$-simplex is added when all directed edges between its $(k+1)$ vertices are present and satisfy the required orientation. This construction respects the directionality of information flow throughout the filtration.

The resulting filtration yields a \textit{persistence diagram} (PD) for each homology dimension $d \geq 0$, which summarizes the birth and death of $d$-dimensional topological features (e.g., connected components, directed cycles, higher-order cavities) as the network is built up from strongest to weakest connections. These diagrams capture the global organization of information transfer in the system at multiple scales and resolutions.

\subsection*{Topological Feature Extraction: Betti Curve}
 
For each homology dimension $d$, we derive a Betti curve $\beta_d(\epsilon)$, which tracks the number of $d$-dimensional features (e.g., connected components for $d{=}0$, loops for $d{=}1$, etc.) present as a function of the filtration threshold $\epsilon$, shown in Equation~\ref{eq:betti-curve}. These Betti curves are interpretable profiles that encapsulate the evolving topological structure of the network across edge-weights. To obtain a compact summary of each Betti curve, we also compute its integral: the Area Under the Betti Curve (AUBC). The AUBC provides a single-number quantification of the total topological activity in dimension $d$ over the entire filtration, Equation~\ref{eq:AUBC}. 

\begin{equation}
  \beta_d(\epsilon) \;=\; \sum_{(b_i,d_i)\in \mathcal{P}_d} \mathbb{I}\!\left[b_i \le \epsilon < d_i\right].
  \label{eq:betti-curve}
\end{equation}

\begin{equation}
    \text{AUBC}_d = \int_{\epsilon_{\min}}^{\epsilon_{\max}} \beta_d(\epsilon) \, d\epsilon
    \label{eq:AUBC}
\end{equation}

This process yields a fixed-length vector of topological summaries per window (or per trial), which can be tracked over time, compared across conditions, or visualized as trajectories in topological space.

\subsection*{Optional: Time-Resolved Analysis}

In experiments involving temporally extended tasks (e.g., behaviorally annotated recordings), we apply the above pipeline in a sliding-window fashion. For each window, the TE matrix, persistence diagrams, and AUBC values are computed independently. Plotting AUBC$_d(t)$ as a function of time reveals the dynamic evolution of topological complexity, enabling temporal alignment with behavioral or cognitive events.

\subsection*{Remarks on Representational Choices}

While we primarily report the Betti Curve and AUBC values in each homology dimension as interpretable and low-dimensional topological summaries, we emphasize that this is a \textit{lossy representation}. The full persistence diagrams contain richer information, including feature lifespans and geometric structure, and may be more suitable for downstream machine learning applications or statistical analyses.

Readers are encouraged to consider alternative vectorizations of persistence diagrams depending on the application, including:
\begin{itemize}
    \item \textbf{Persistence images} \citep{adams2017persistence}: kernel-based density estimators over PDs.
    \item \textbf{Persistence landscapes} \citep{bubenik2015statistical}: functional representations supporting statistical testing.
    \item \textbf{Persistence entropy}~\citep{CHINTAKUNTA2015391} and \textbf{silhouettes}~\citep{chazai2014sil}: scalar or weighted summaries of PDs.
\end{itemize}

Our use of Betti Curves focuses on providing an accessible, task-discriminative representation that supports comparisons across tasks, time, and experimental conditions, while maintaining interpretability of topological activity in each dimension. 

\section{Experiment 1: Synthetic Task: Logic Gates}

This experiment investigates the topological structure that emerges when a spiking neural network (SNN) processes binary logic operations. The task was deliberately designed to be trivial yet systematically varied in complexity, serving as a controlled benchmark for analyzing emerging topological patterns. By selecting a fully observable task domain with well-defined input–output mappings, we aim to isolate and demonstrate topological emergence without confounding factors. This provides a foundation for applying the pipeline to more complex or real-world datasets.

The network is trained to perform three Boolean functions: \texttt{AND}, \texttt{OR}, and \texttt{XOR}, over randomly sampled 2-bit binary inputs. Each trial consists of a binary input vector $x \in \{0,1\}^2$ and a one-hot encoded gate selector $g \in \{(1,0,0), (0,1,0), (0,0,1)\}$ corresponding to \texttt{AND}, \texttt{OR}, or \texttt{XOR}, respectively. These are concatenated into a 5-dimensional vector, then converted into Poisson spike trains over a 20\,ms simulation window using high (0.5) and low (0.02) firing rates for logical 1 and 0, respectively.

A fixed recurrent spiking network of 64 leaky integrate-and-fire (LIF) neurons processes the inputs. After training, a random test set of 4096 examples is evaluated with frozen network weights. To isolate the effect of the logic task on the resulting topological structure:
\begin{itemize}
    \item All inputs include exactly one active gate selector, ensuring that input energy remains constant across tasks.
    \item The same network is used for all gates, allowing us to attribute differences in emergent topology solely to the gate logic and resulting dynamics.
\end{itemize}

\subsection{Topological Pipeline and Analysis}

For each test sample, we compute pairwise transfer entropy (TE) between neurons over the 20\,ms window, construct a weighted directed graph, and apply a sublevel filtration using the TE edge-weight values. Because lower values must correspond to stronger connections under a sublevel filtration, TE values are inverted via a monotonic transformation (mirrored about zero), so that edges with higher information transfer appear earlier in the filtration (see Section~\ref{sec:pipeline}). Persistent homology is computed on the resulting directed flag complex using the \texttt{Flagser} library.

For each homology dimension $d \in [0,7]$, we first analyse the class-mean \emph{Betti curve}, which profiles the number of $d$-dimensional features across the filtration. We then summarise each trial’s curve by its \emph{area under the Betti curve} (AUBC), yielding a scalar descriptor of topological activity. %

\subsection{Results: Emergent High-Dimensional Structure of XOR}

\begin{figure}[ht]
    \centering
    \begin{subfigure}[t]{0.48\linewidth}
        \includegraphics[width=\linewidth]{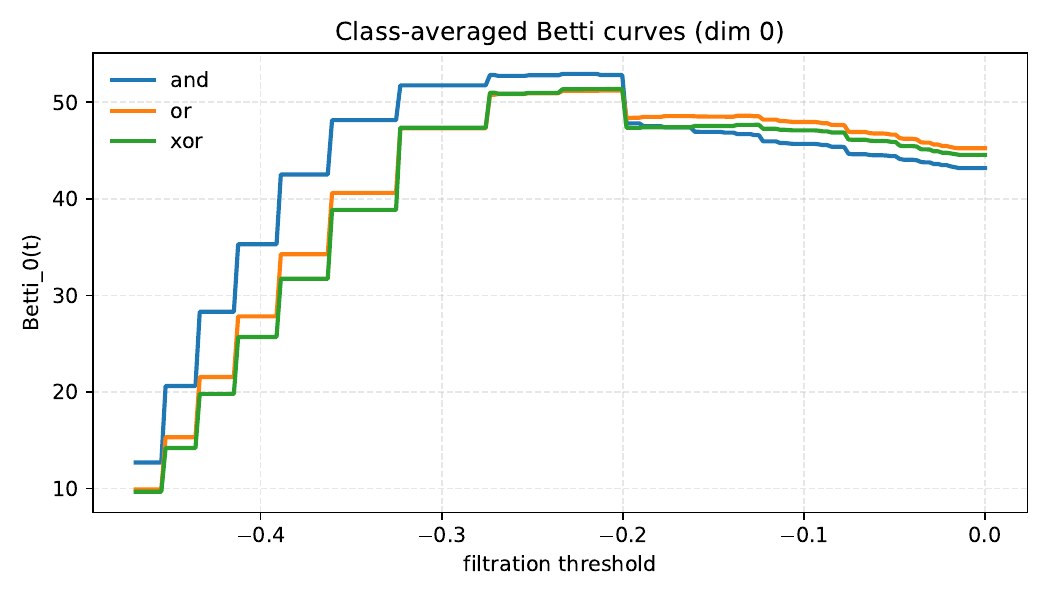}
    \end{subfigure}\hfill
    \begin{subfigure}[t]{0.48\linewidth}
        \includegraphics[width=\linewidth]{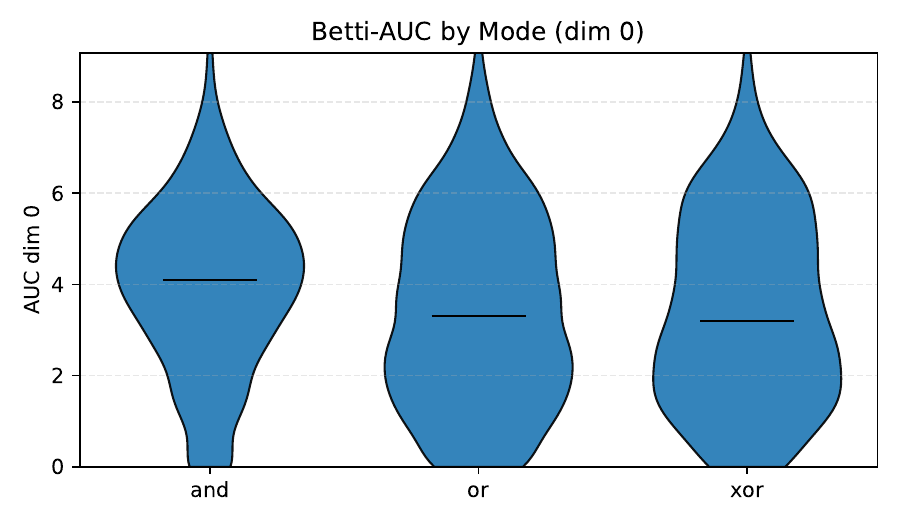}
    \end{subfigure}

    \vspace{0.6em}

    \begin{subfigure}[t]{0.48\linewidth}
        \includegraphics[width=\linewidth]{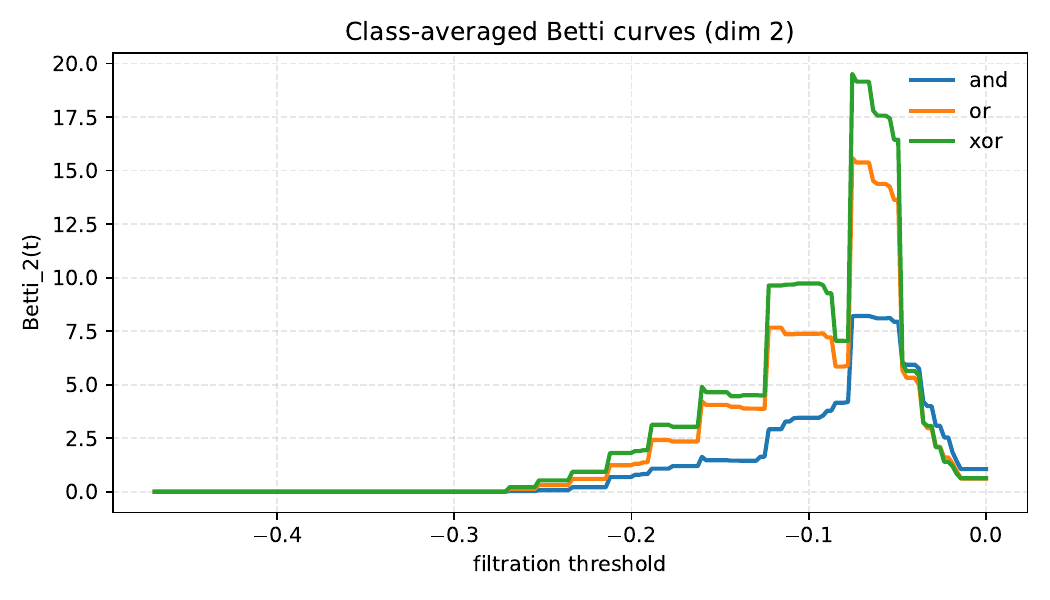}
    \end{subfigure}\hfill
    \begin{subfigure}[t]{0.48\linewidth}
        \includegraphics[width=\linewidth]{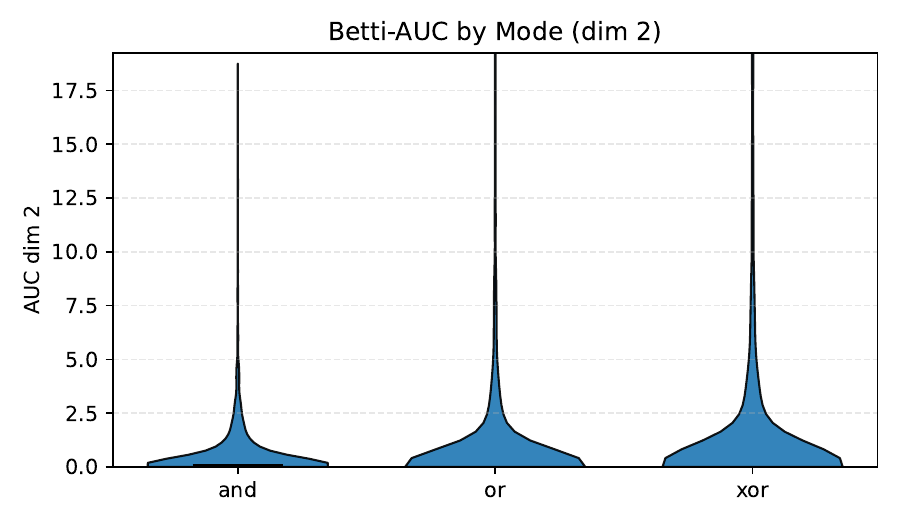}
    \end{subfigure}

    \vspace{0.6em}

    \begin{subfigure}[t]{0.48\linewidth}
        \includegraphics[width=\linewidth]{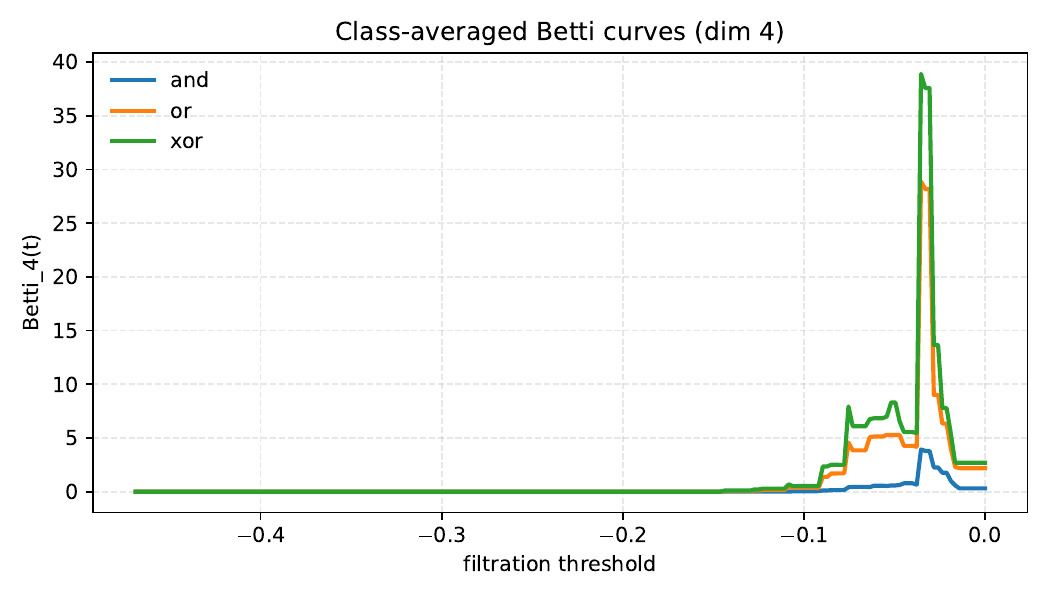}
    \end{subfigure}\hfill
    \begin{subfigure}[t]{0.48\linewidth}
        \includegraphics[width=\linewidth]{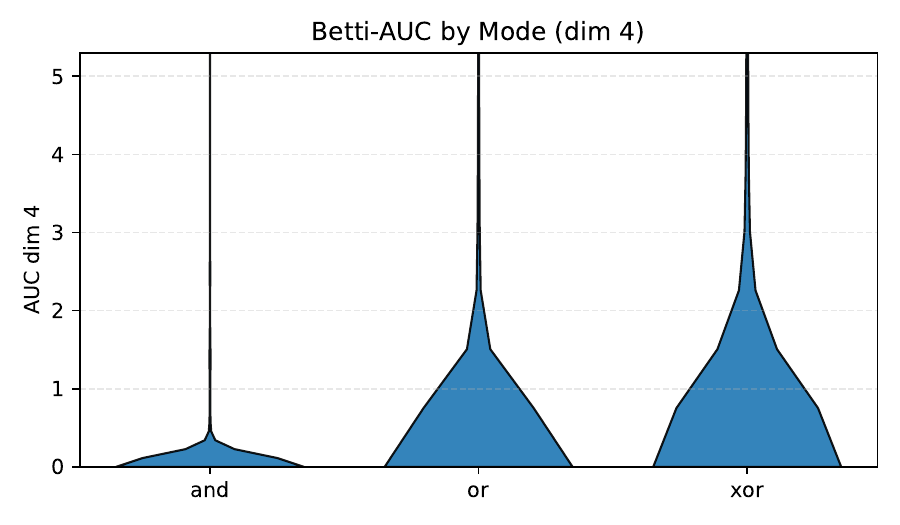}
    \end{subfigure}

    \vspace{0.6em}

    \begin{subfigure}[t]{0.48\linewidth}
        \includegraphics[width=\linewidth]{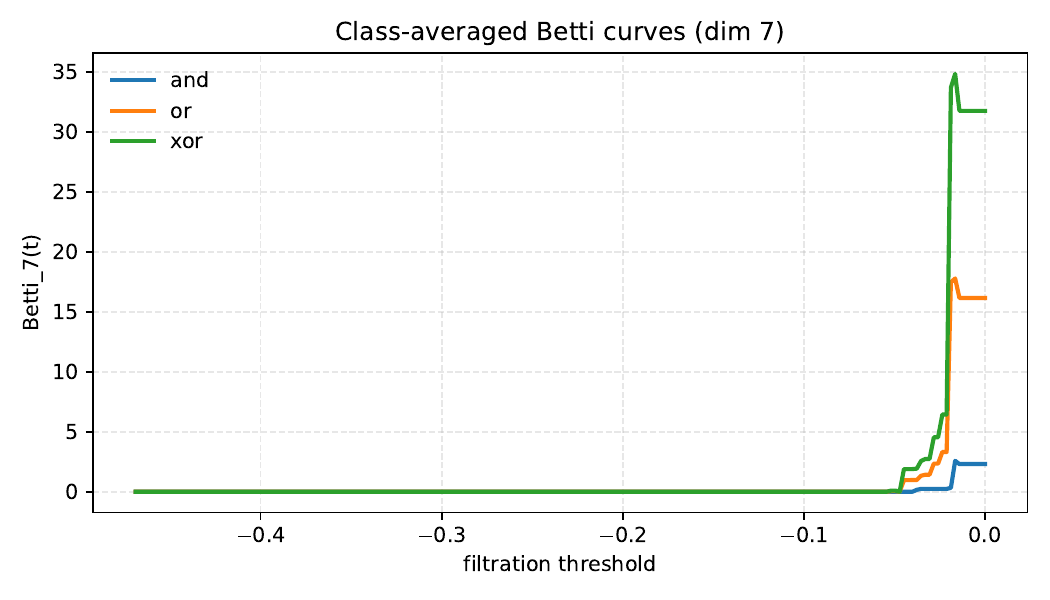}
    \end{subfigure}\hfill
    \begin{subfigure}[t]{0.48\linewidth}
        \includegraphics[width=\linewidth]{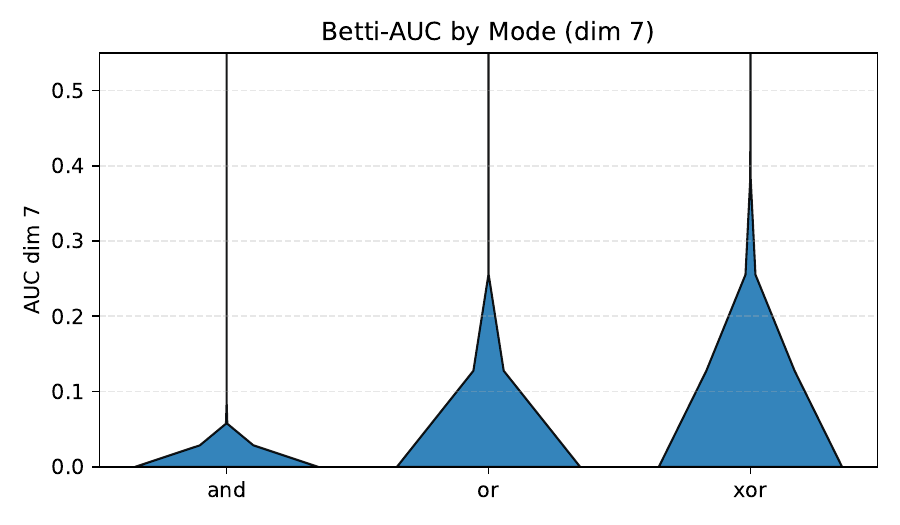}
    \end{subfigure}

    \caption{Logic gates: \textbf{Betti curves (left)} and corresponding \textbf{AUBC distributions (right)} for representative homology dimensions ($d=0,2,4,7$). Curves show class-mean Betti profiles for \texttt{AND}, \texttt{OR}, and \texttt{XOR} across 4096 test samples; violins summarise per-trial AUBC for the same dimension. Filtration proceeds left-to-right; lower (more negative) edge weights represent stronger TE (Section~\ref{sec:pipeline}).}
    \label{fig:logic_bc_aubc}
\end{figure}

To visualise how information-flow topology differs by logic operation, we first examine the Betti curves for each gate. Figure~\ref{fig:logic_bc_aubc} (left column) shows the class-mean Betti curves for representative homology dimensions ($d=0,2,4,7$). In low dimensions like $d=0$, the curves for all three gates closely overlap, indicating similar counts of connected components across the filtration. As the homology dimension increases, clear differences emerge: by $d=4$ and $d=7$, the curve for \texttt{XOR} lies consistently above those for \texttt{AND} and \texttt{OR} across a broad range of thresholds. This higher Betti-curve amplitude means \texttt{XOR} sustains more topological features (loops and higher-order cavities) through the filtration, reflecting its greater computational complexity.

To quantify these curve-level differences, we compute the AUBC for every trial and dimension. The right column of Figure~\ref{fig:logic_bc_aubc} presents the resulting AUBC distributions. Consistent with the Betti profiles, \texttt{XOR} shows both higher mean AUBC and greater variance than \texttt{AND}/\texttt{OR} in higher dimensions, while all three tasks remain similar in low dimensions.

\paragraph{High-dimensional divergence}
In lower dimensions, all three gates produce similar topological activity. However, for $d \geq 2$ (illustrated at $d=2,4,7$ in Figure~\ref{fig:logic_bc_aubc}), \texttt{XOR} exhibits both:
\begin{itemize}
    \item higher average AUBC, indicating greater overall topological richness; and
    \item larger variance, suggesting more diverse topological configurations across samples.
\end{itemize}
In other words, \texttt{XOR} not only increases the total number of persistent features (higher mean AUBC) but also induces more trial-to-trial variability in the emergent topology compared to the simpler gates.

\section{Experiment 2: Synthetic Task: Image Classification}

To complement the logic gate experiment’s categorical complexity, this experiment explores how increasing perceptual difficulty, introduced via structured noise, affects the emergent topologies of a spiking network. We use the MNIST dataset, a widely used benchmark in computer vision consisting of 28$\times$28 grayscale images of handwritten digits (0 through 9). Each image is treated as a 2D pixel array with values in $[0, 1]$, representing background and ink intensity.

We conduct two experiments using different noise schemes, Section~\nameref{sec:mnist:flip} and Section~\nameref{sec:mnist:move}. These experiments allow us to probe how the topological structures of network activity evolve with quantitative complexity, while controlling for input structure and (in the second experiment) input energy.

\subsection{Network and Encoding Details}

Each image is flattened and encoded into spike trains using Poisson rate coding: pixel values are converted to spike rates, with 0.5 representing white (active) and 0.02 for black (inactive). These spike trains are presented to a spiking network with 64 leaky integrate-and-fire (LIF) neurons over a 100\,ms time window. A readout layer maps the network dynamics to digit classes.

Each experiment uses a single network trained on a mixture of all noise levels and evaluated in a frozen state. Persistent homology is computed on the Transfer Entropy (TE) graphs derived from spiking activity, following the pipeline described previously (Section~\ref{sec:pipeline}): we analyze class-mean \emph{Betti curves} per homology dimension and summarize each trial with the \emph{area under the Betti curve} (AUBC). %

\subsection{Sub-Experiment A: Bit-Flipping Noise} \label{sec:mnist:flip}

To inject structured complexity, we randomly flip a fixed number of pixels in each image (changing 0s to 1s or vice versa). Because MNIST images contain predominantly black background pixels, this process tends to increase the number of white pixels (i.e., spiking inputs), introducing both perceptual distortion and greater network stimulation.

\begin{figure}[t]
  \centering
  \begin{subfigure}{0.25\textwidth}
    \centering
    \includegraphics[width=\linewidth]{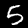}
    \subcaption{N0}\label{fig:MNIST_flip:N0}
  \end{subfigure}\hfill
  \begin{subfigure}{0.25\textwidth}
    \centering
    \includegraphics[width=\linewidth]{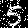}
    \subcaption{N4}\label{fig:MNIST_flip:N4}
  \end{subfigure}\hfill
  \begin{subfigure}{0.25\textwidth}
    \centering
    \includegraphics[width=\linewidth]{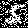}
    \subcaption{N8}\label{fig:MNIST_flip:N8}
  \end{subfigure}
  \caption{Three images from the MNIST dataset with applied bit-flipping noise. Examples represent noise levels 0, 4, and 8. The noise increases linearly across levels.}
  \label{fig:MNIST_flip}
\end{figure}

Noise levels for bit-flipping can be seen in Figure~\ref{fig:MNIST_flip}. Flip counts per noise level are defined in Table~\ref{tab:flip_noise_levels}:
\begin{table}[t]
    \centering
    \small
    \caption{Noise levels for bit-flipping. Columns represent noise levels \texttt{n0}--\texttt{n8}; values indicate the number of random pixel flips applied to each image.}
    \label{tab:flip_noise_levels}
    \begin{tabular}{ccccccccc}
        \hline
        \texttt{n0} & \texttt{n1} & \texttt{n2} & \texttt{n3} & \texttt{n4} & \texttt{n5} & \texttt{n6} & \texttt{n7} & \texttt{n8} \\
        \hline
        0 & 25 & 50 & 75 & 100 & 125 & 150 & 175 & 200 \\
        \hline
    \end{tabular}
\end{table}

\begin{figure}[ht]
    \centering
    \begin{subfigure}[t]{0.48\linewidth}
        \includegraphics[width=\linewidth]{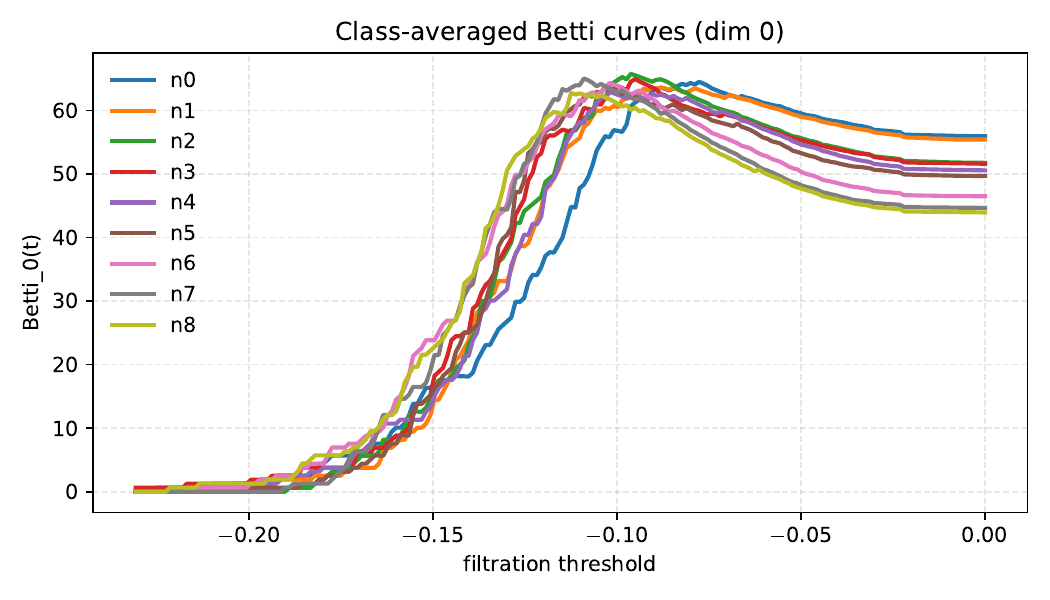}
    \end{subfigure}\hfill
    \begin{subfigure}[t]{0.48\linewidth}
        \includegraphics[width=\linewidth]{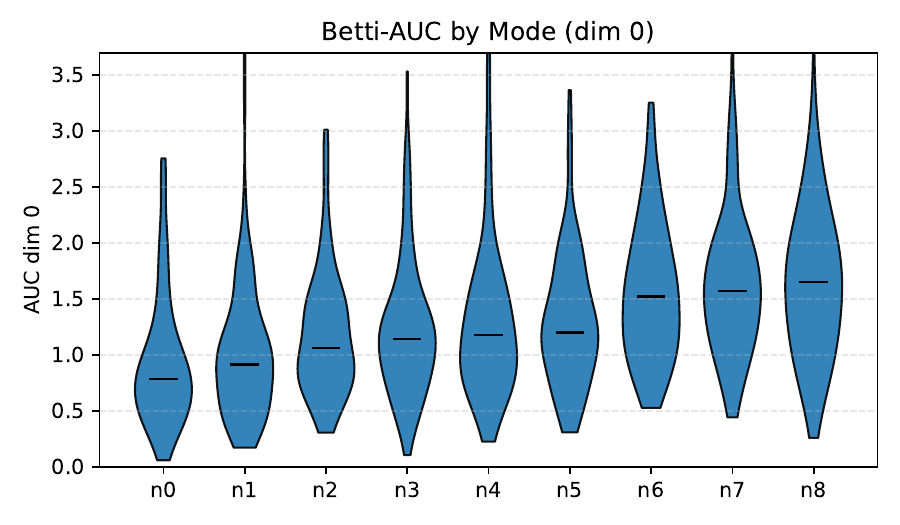}
    \end{subfigure}

    \vspace{0.6em}

    \begin{subfigure}[t]{0.48\linewidth}
        \includegraphics[width=\linewidth]{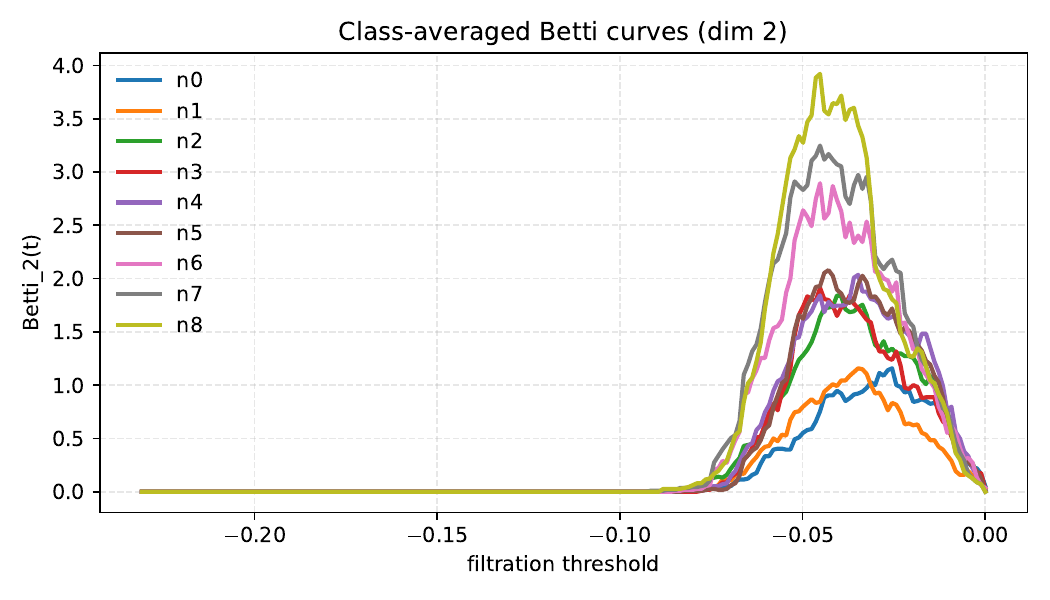}
    \end{subfigure}\hfill
    \begin{subfigure}[t]{0.48\linewidth}
        \includegraphics[width=\linewidth]{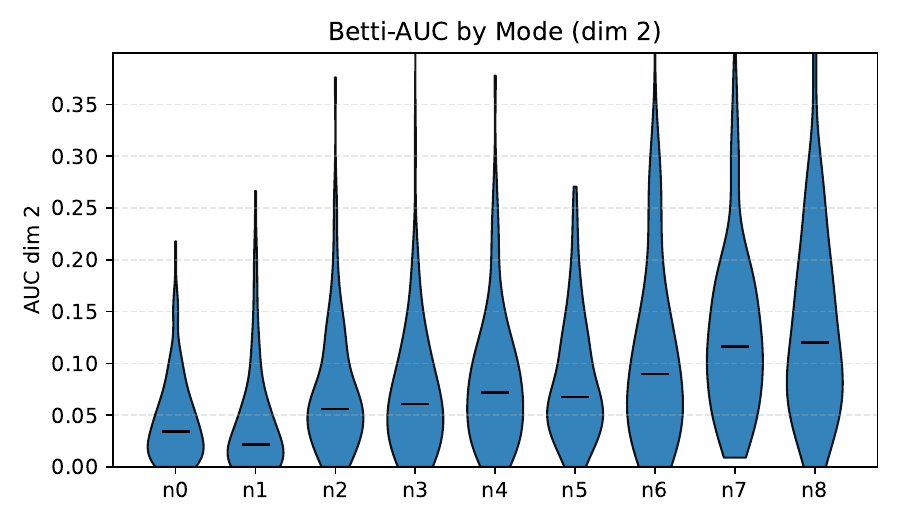}
    \end{subfigure}

    \vspace{0.6em}

    \begin{subfigure}[t]{0.48\linewidth}
        \includegraphics[width=\linewidth]{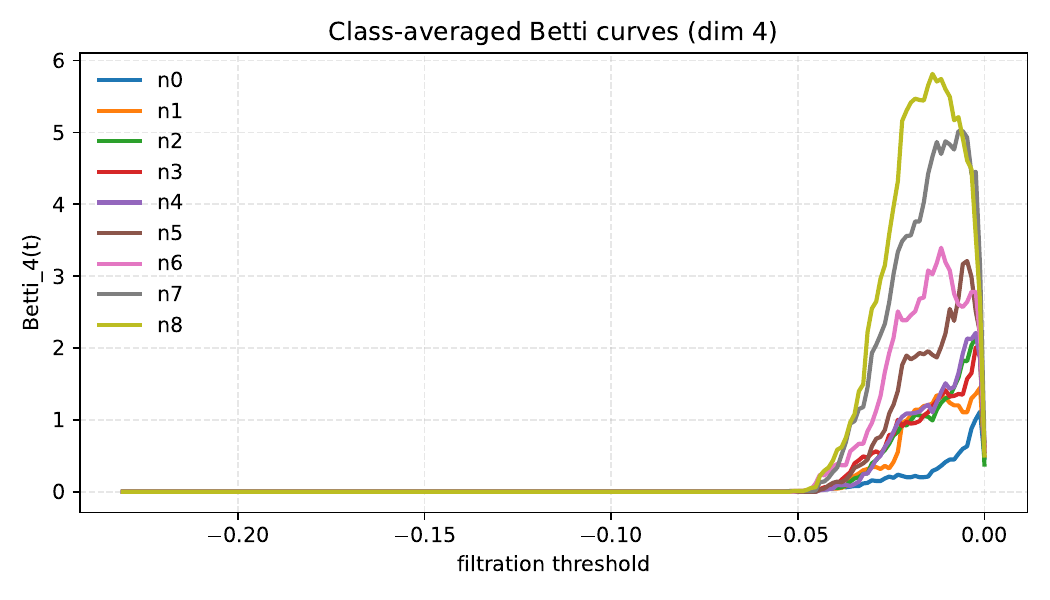}
    \end{subfigure}\hfill
    \begin{subfigure}[t]{0.48\linewidth}
        \includegraphics[width=\linewidth]{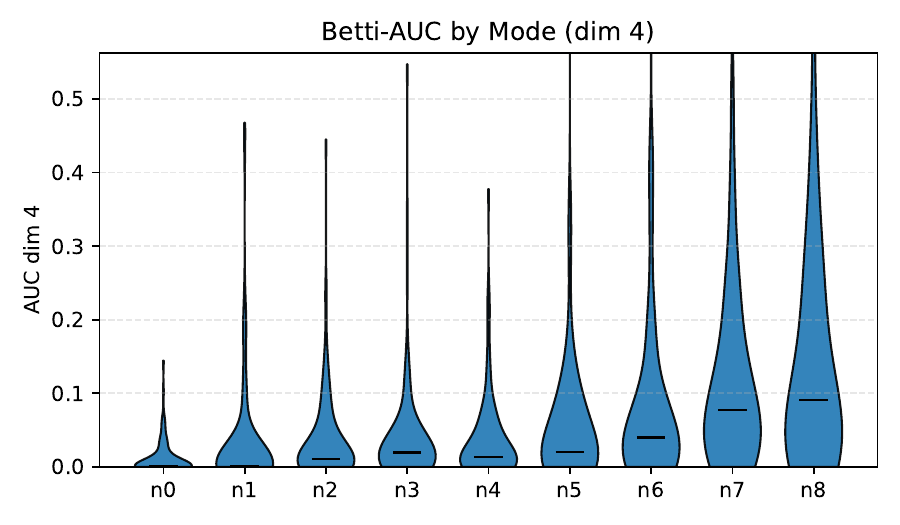}
    \end{subfigure}

    \vspace{0.6em}

    \begin{subfigure}[t]{0.48\linewidth}
        \includegraphics[width=\linewidth]{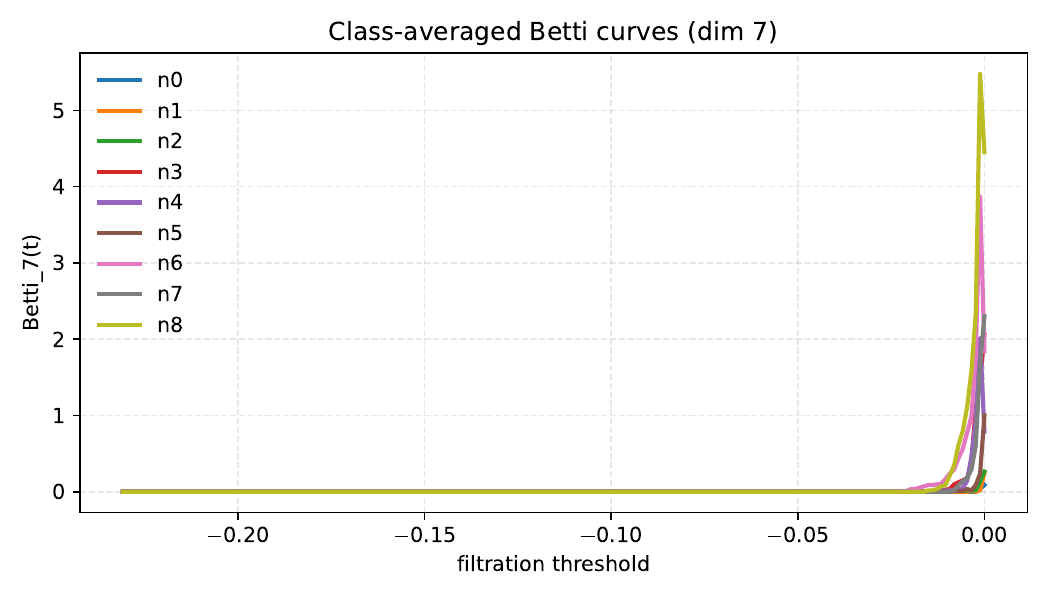}
    \end{subfigure}\hfill
    \begin{subfigure}[t]{0.48\linewidth}
        \includegraphics[width=\linewidth]{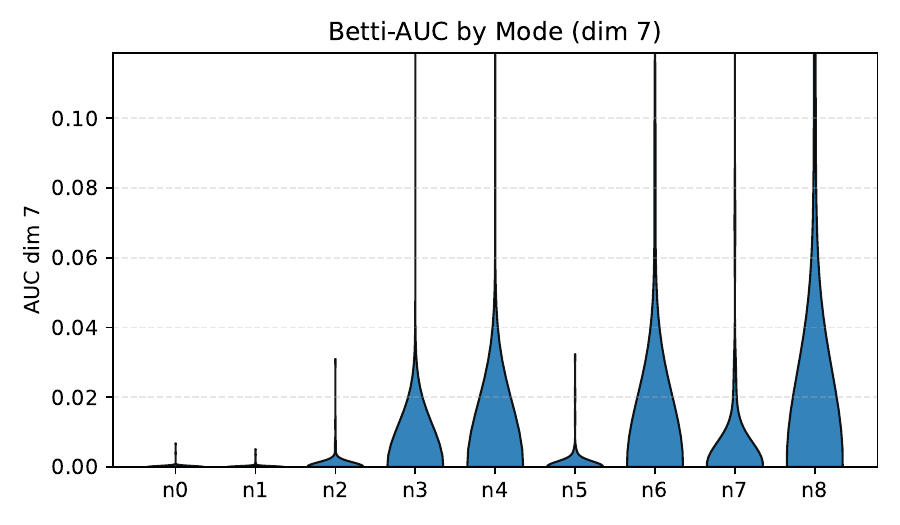}
    \end{subfigure}

    \caption{MNIST bit-flipping: \textbf{Betti curves (left)} and corresponding \textbf{AUBC distributions (right)} for representative homology dimensions ($d=0,2,4,7$). Curves show class-mean Betti profiles per noise level (\texttt{n0}--\texttt{n8}); violins summarise per-trial AUBC for the same dimension. Filtration proceeds left-to-right; lower (more negative) edge weights represent stronger TE (Section~\ref{sec:pipeline}).}
    \label{fig:mnist_flip_bc_aubc}
\end{figure}

\paragraph{Betti curves and AUBC.}
Figure~\ref{fig:mnist_flip_bc_aubc} merges the class-mean Betti curves (left) with the corresponding AUBC distributions (right) for representative homology dimensions ($d=0,2,4,7$). In $d=0$, Betti curves are similar across noise conditions, consistent with comparable connected-component counts early in the filtration. As $d$ increases, curves separate: higher noise levels maintain elevated Betti curves across a wider threshold range, indicating richer topological structure. Quantitatively, the AUBC distributions mirror this pattern: means and variances grow with noise, especially for $d\in\{4,7\}$.

\clearpage
\subsection{Sub-Experiment B: Bit-Moving Noise (Pixel-Preserving)}\label{sec:mnist:move}

To isolate the effects of perceptual distortion from changes in input quantity, we perform pixel-preserving perturbations. In this scheme, a white (foreground) pixel is removed and placed at a randomly selected black (background) location. This transformation maintains the same number of white pixels, and therefore the same overall spike energy and image resolution.

\begin{figure}[t]
  \centering
  \begin{subfigure}{0.25\textwidth}
    \centering
    \includegraphics[width=\linewidth]{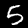}
    \subcaption{N0}\label{fig:MNIST_move:N0}
  \end{subfigure}\hfill
  \begin{subfigure}{0.25\textwidth}
    \centering
    \includegraphics[width=\linewidth]{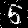}
    \subcaption{N4}\label{fig:MNIST_move:N4}
  \end{subfigure}\hfill
  \begin{subfigure}{0.25\textwidth}
    \centering
    \includegraphics[width=\linewidth]{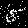}
    \subcaption{N8}\label{fig:MNIST_move:N8}
  \end{subfigure}
  \caption{Three images from the MNIST dataset with applied bit-moving noise. Examples represent noise levels 0, 4, and 8. The noise increases linearly across levels.}
  \label{fig:MNIST_move}
\end{figure}

Noise levels for bit-moving can be seen in Figure~\ref{fig:MNIST_move}. Pixel-moves per level are defined in Table~\ref{tab:move_noise_levels}:
\begin{table}[t]
    \centering
    \small
    \caption{Noise levels for pixel-moving. Columns represent noise levels \texttt{n0}--\texttt{n8}; values indicate the number of random pixel movements per image, relocating white foreground pixels to black background locations.}
    \label{tab:move_noise_levels}
    \begin{tabular}{ccccccccc}
        \hline
        \texttt{n0} & \texttt{n1} & \texttt{n2} & \texttt{n3} & \texttt{n4} & \texttt{n5} & \texttt{n6} & \texttt{n7} & \texttt{n8} \\
        \hline
        0 & 5 & 10 & 15 & 20 & 25 & 30 & 35 & 40 \\
        \hline
    \end{tabular}
\end{table}

\begin{figure}[ht]
    \centering
    \begin{subfigure}[t]{0.48\linewidth}
        \includegraphics[width=\linewidth]{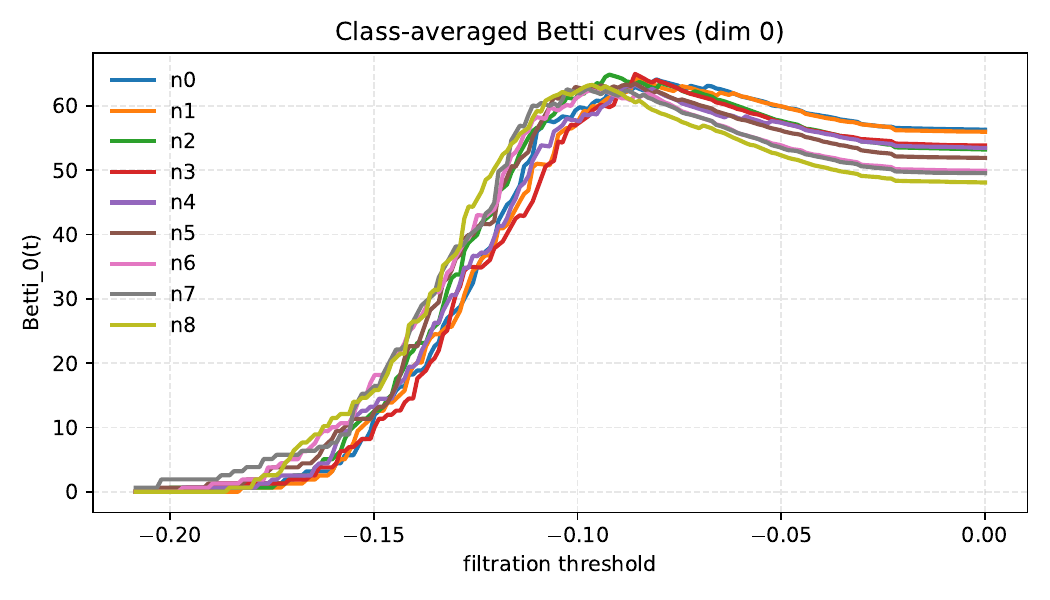}
    \end{subfigure}\hfill
    \begin{subfigure}[t]{0.48\linewidth}
        \includegraphics[width=\linewidth]{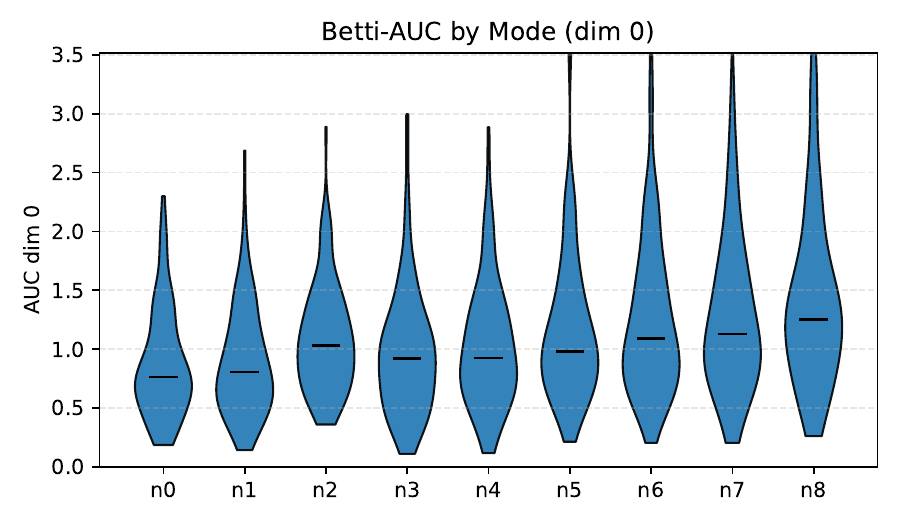}
    \end{subfigure}

    \vspace{0.6em}

    \begin{subfigure}[t]{0.48\linewidth}
        \includegraphics[width=\linewidth]{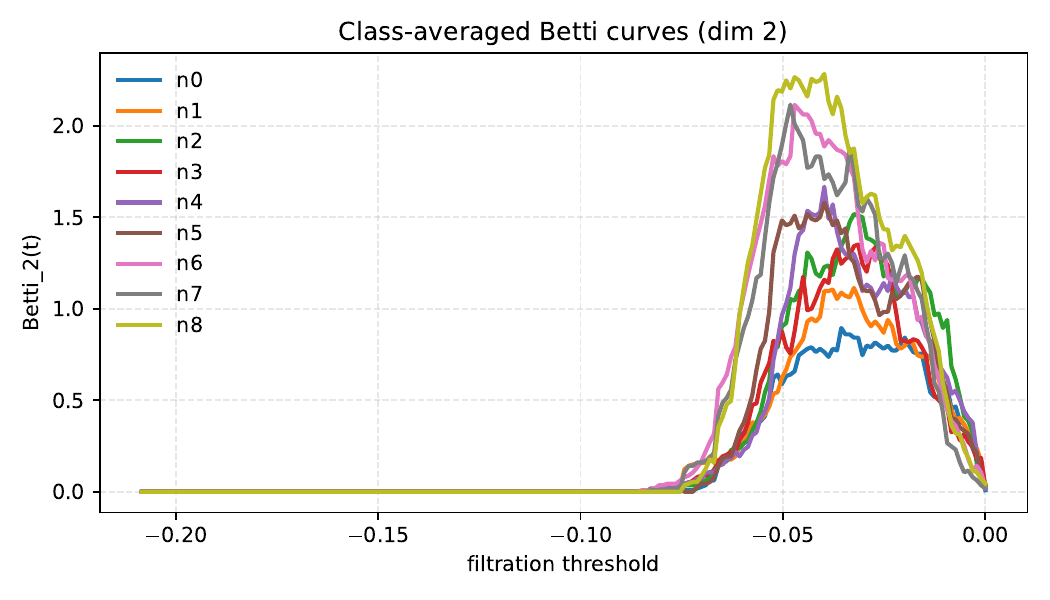}
    \end{subfigure}\hfill
    \begin{subfigure}[t]{0.48\linewidth}
        \includegraphics[width=\linewidth]{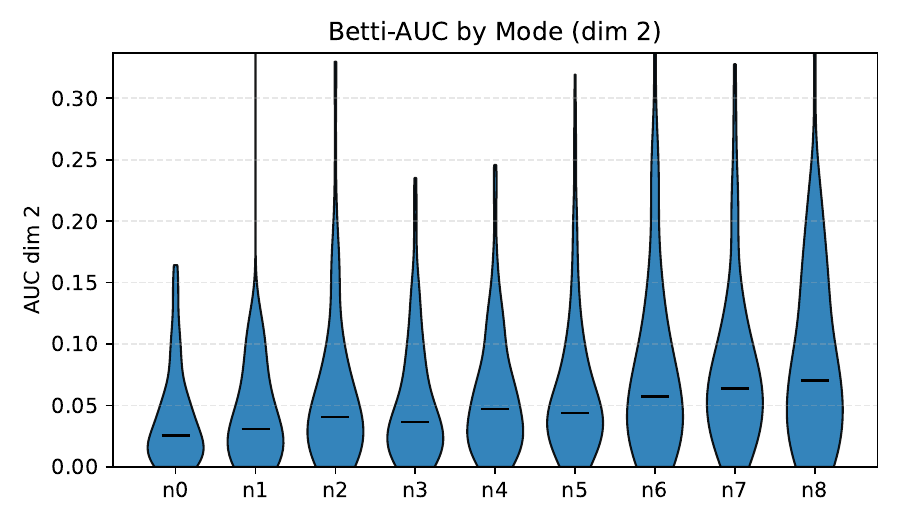}
    \end{subfigure}

    \vspace{0.6em}

    \begin{subfigure}[t]{0.48\linewidth}
        \includegraphics[width=\linewidth]{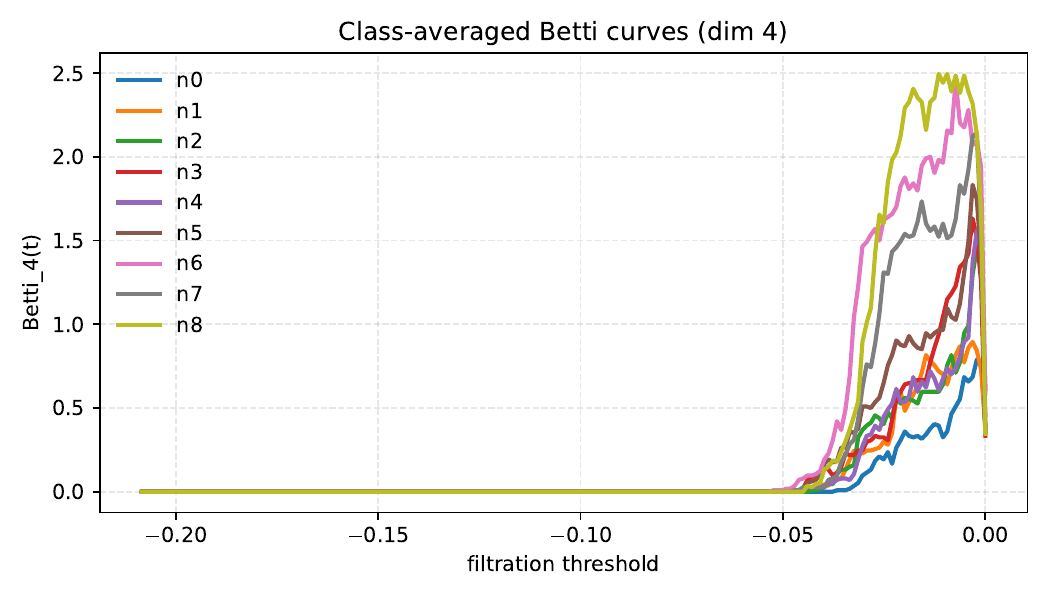}
    \end{subfigure}\hfill
    \begin{subfigure}[t]{0.48\linewidth}
        \includegraphics[width=\linewidth]{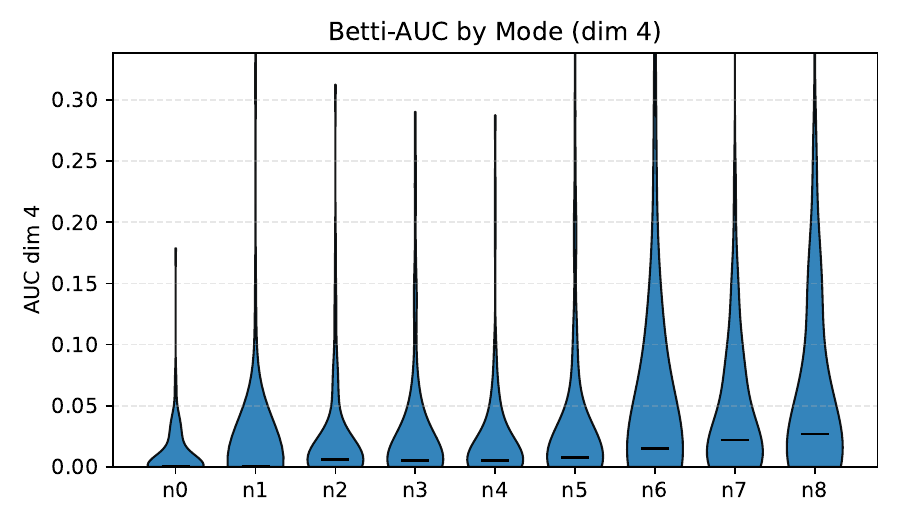}
    \end{subfigure}

    \vspace{0.6em}

    \begin{subfigure}[t]{0.48\linewidth}
        \includegraphics[width=\linewidth]{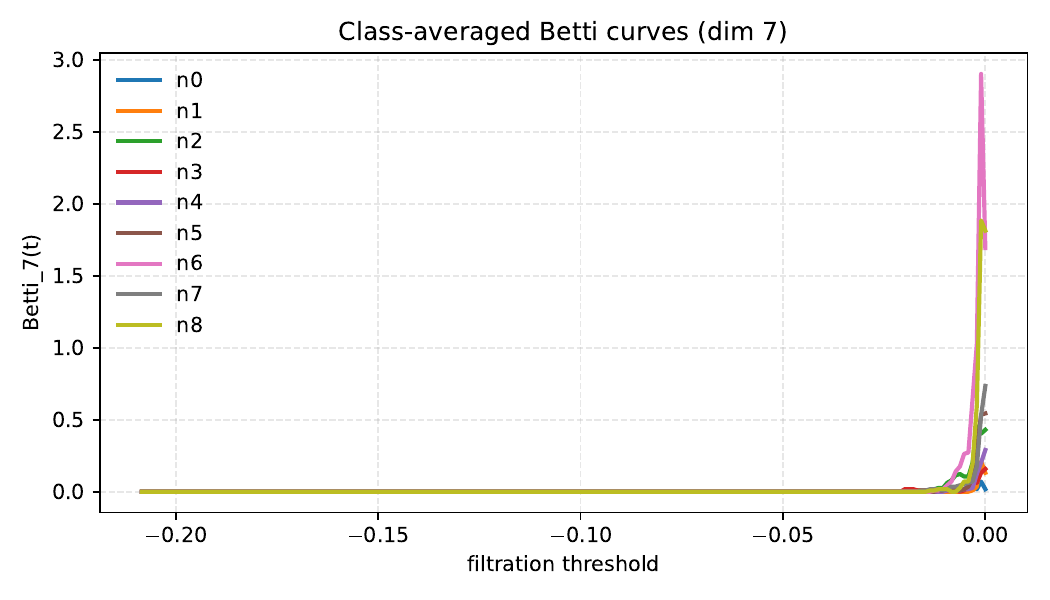}
    \end{subfigure}\hfill
    \begin{subfigure}[t]{0.48\linewidth}
        \includegraphics[width=\linewidth]{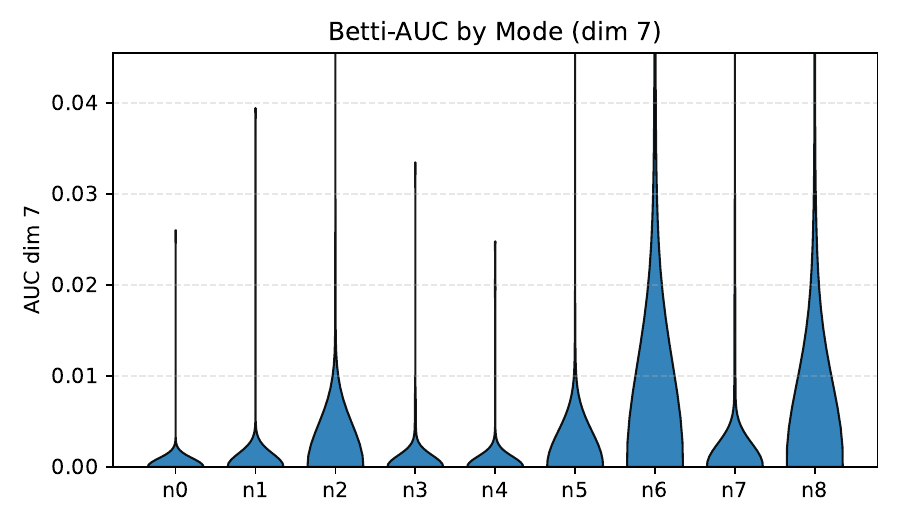}
    \end{subfigure}

    \caption{MNIST bit-moving (pixel-preserving): \textbf{Betti curves (left)} and corresponding \textbf{AUBC distributions (right)} for representative homology dimensions ($d=0,2,4,7$). Curves show class-mean Betti profiles per noise level (\texttt{n0}--\texttt{n8}); violins summarise per-trial AUBC for the same dimension. Filtration proceeds left-to-right; lower (more negative) edge weights represent stronger TE (Section~\ref{sec:pipeline}).}
    \label{fig:mnist_move_bc_aubc}
\end{figure}

\paragraph{Betti curves and AUBC}
Figure~\ref{fig:mnist_move_bc_aubc} shows the merged Betti curves (left) and AUBC distributions (right) for $d=0,2,4,7$. Because total spike energy is preserved across noise levels, the $d=0$ curves remain closely aligned. Nevertheless, as $d$ increases, higher noise levels yield consistently elevated Betti curves sustained over broader threshold ranges, indicating more persistent higher-order structure due purely to geometric rearrangement. The AUBC distributions reflect this: both means and variances rise with noise, most notably for $d\in\{4,7\}$, even though total input energy is fixed.

\subsection{Interpretation}

Across both noise paradigms, Betti curves are smooth and display systematic shifts with increasing perceptual difficulty, particularly in higher homology dimensions. The AUBC evolution quantitatively matches these curve-level differences: means and variances grow with noise, reflecting richer and more variable topological organization. Crucially, the pixel-moving results demonstrate that these effects persist even when total input energy is held constant, indicating that topological fingerprints are sensitive to geometry-driven changes in information flow rather than merely input magnitude.

\section{Experiment 3: Biological Mice Data}

This experiment examines topological structure in neural spike data from the International Brain Laboratory (IBL) Neuropixels project~\citep{ibl2021neuro}. To isolate topological changes from confounds like anatomy or sensor placement, all trials were drawn from a single subject (\texttt{ZM\_2240}) and a fixed set of 256 putative neurons from one probe.

Each trial involved a head-fixed mouse performing a two-alternative forced choice task using a rotating wheel. Visual stimuli were presented, after which the mouse made a movement decision. If correct, feedback was provided via auditory cue and reward.

\subsection{Topological Analysis Pipeline}
Neural recordings were sampled at 1\,kHz. Each trial was segmented using a 100\,ms sliding window with a 25\,ms step size. Persistent homology was computed for each window, yielding a sequence of AUBC values across Betti dimensions $d = 0,1,2,3$.

We aligned the time windows to five task-defined behavioral epochs:
\begin{itemize}
    \item \textbf{Baseline:} From trial start up to stimulus onset.
    \item \textbf{Stimon:} 100\,ms after stimulus onset.
    \item \textbf{Trial:} From 100\,ms post-stimulus to feedback onset.
    \item \textbf{Feedback:} 100\,ms after feedback is delivered.
    \item \textbf{Post-feedback:} From 100\,ms after feedback until trial end.
\end{itemize}

\subsection{Topological Metrics}

Table~\ref{tab:mouse_metrics} summarizes the metrics used to quantify topological change per dimension and behavioral window:

\begin{table}[H]
\centering
\caption{Topological metrics used to evaluate behavioral modulation of persistence.}
\label{tab:mouse_metrics}
\renewcommand{\arraystretch}{1.5}
\begin{tabular}{p{3.5cm} | p{5.5cm} | p{6cm}}
\toprule
\textbf{Metric} & \textbf{Equation} & \textbf{Description} \\
\midrule
Mean AUBC & $\mu_{i,w}^{(d)}$ & Average AUBC in trial $i$, dimension $d$, and window $w$. \\
\midrule
Delta to Baseline & $\Delta_{i,w}^{(d)} = \mu_{i,w}^{(d)} - \mu_{i,\text{baseline}}^{(d)}$ & Absolute increase in topological activity relative to pre-stimulus baseline. \\
\midrule
Ratio to Baseline & $R_{i,w}^{(d)} = \frac{\mu_{i,w}^{(d)}}{\mu_{i,\text{baseline}}^{(d)}}$ & Relative change in persistence compared to baseline. Ratio $>1$ implies increase. \\
\midrule
Cohen’s $d$ & $d_w^{(d)} = \frac{\bar{\mu}_w^{(d)} - \bar{\mu}_b^{(d)}}{s_{\text{pooled}}}$ & Effect size comparing mean AUBC in window $w$ vs. baseline. Pooled standard deviation is:
\\[0.5em]
& $s_{\text{pooled}} = \sqrt{\frac{\sigma_b^2 + \sigma_w^2}{2}}$ & Combines variability across both conditions. \\
\midrule
Mutual \newline Information & $I(X^{(d)}; Y) \newline = \sum_{x, y} P(x, y) \log \left( \frac{P(x, y)}{P(x) P(y)} \right)$ & Captures nonlinear dependence between AUBC values ($X^{(d)}$) and behavioral window labels ($Y$). \\
\bottomrule
\end{tabular}
\end{table}

\subsection{Results and Interpretation}
\begin{figure}[H]
    \centering
    \includegraphics[width=0.9\linewidth]{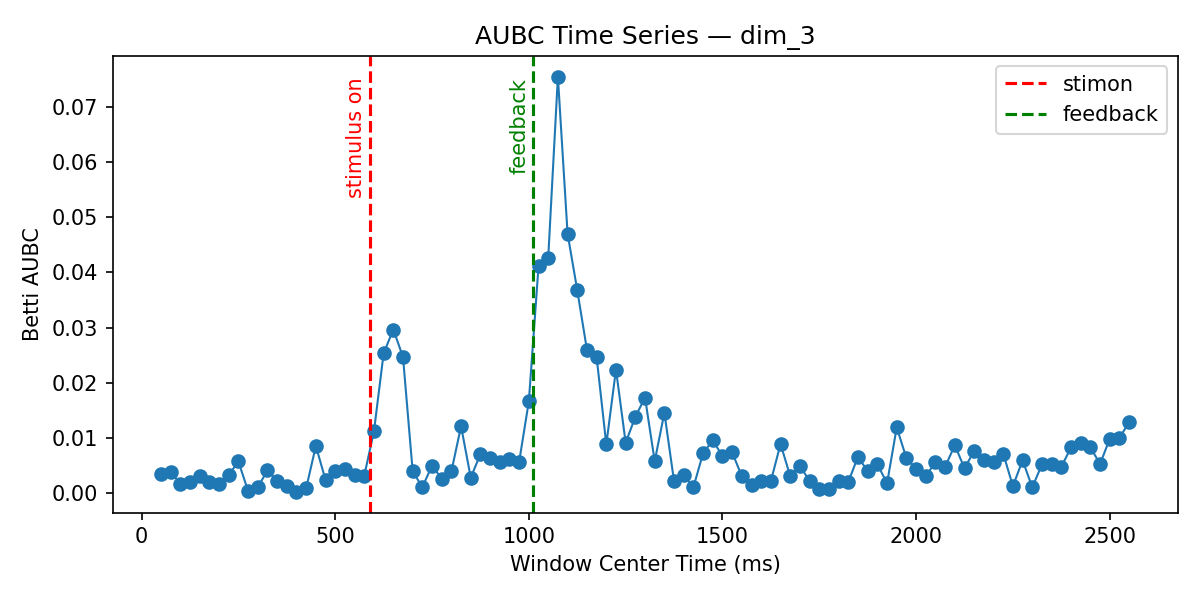}
    \caption{AUBC time series for Betti-3 during a single trial. Vertical lines indicate stimulus and feedback onset.}
    \label{fig:mouse_aubc_timeseries}
\end{figure}

Figure~\ref{fig:mouse_aubc_timeseries} shows a sample trial, plotting dimension-3 AUBC values over time alongside event-aligned behavioral windows.

Table~\ref{tab:mouse_summary} reports the averaged topological metrics across 153 trials. Dimension 1 shows the most salient modulation: during feedback, it reached a ratio of 17.73, a delta of 10.25, and a Cohen’s $d$ of 5.27, indicating a robust and consistent elevation relative to baseline. This is further supported by its high mutual information (MI = 0.86), implying that topological structure in dimension 1 is strongly informative about behavioral phase.

Higher Betti dimensions (2 and 3) also revealed pronounced responses, especially during feedback. For example, dimension 2 exhibited a 20.59× increase from baseline despite low absolute magnitude, and an effect size of $d=1.55$, again reflected in its MI of 0.72. These findings highlight that higher-dimensional topological features, though sparse, can be meaningful markers of dynamic neural events.

Dimension 0, representing basic connectivity, showed more conservative but reliable increases, with its strongest effect again appearing during feedback. Its mutual information of 0.47 indicates consistent but lower behavioral specificity.

\begin{table}[H]
\centering
\caption{Topological response metrics per behavioral window for each Betti dimension (153 trials, subject \texttt{ZM\_2240}).}
\label{tab:mouse_summary}
\renewcommand{\arraystretch}{1.3}
\begin{tabular}{c|c|cccc}
\toprule
\textbf{Dim} & \textbf{Window} & $\Delta$ & Ratio & Cohen's $d$ & MI \\
\midrule
\multirow{4}{*}{0} & Stimon          & 0.5886 & 1.3187  & 0.7311  & \multirow{4}{*}{0.4708} \\
                   & Trial           & 0.5300 & 1.2919  & 1.0171  &                          \\
                   & Feedback        & 3.1026 & 2.6833  & 2.4185  &                          \\
\midrule
\multirow{3}{*}{1} & Stimon          & 2.8507 & 5.5762  & 2.0421  & \multirow{4}{*}{0.8608} \\
                   & Trial           & 2.2119 & 4.6141  & 1.9318  &                          \\
                   & Feedback        & 10.2538& 17.7306 & 5.2730  &                          \\
\midrule
\multirow{3}{*}{2} & Stimon          & 0.0043 & 3.1970  & 1.2156  & \multirow{4}{*}{0.7173} \\
                   & Trial           & 0.0030 & 2.6611  & 1.0925  &                          \\
                   & Feedback        & 0.0385 & 20.5923 & 1.5540  &                          \\
\midrule
\multirow{3}{*}{3} & Stimon          & 0.0036 & 2.3868  & 0.9610  & \multirow{4}{*}{0.6878} \\
                   & Trial           & 0.0029 & 2.2243  & 1.3205  &                          \\
                   & Feedback        & 0.0255 & 11.1090 & 2.9521  &                          \\
\bottomrule
\end{tabular}
\end{table}

\subsection{Contrast-Conditioned Analysis}
\label{sec:contrast_analysis}

In the IBL task, the visual stimulus on each trial is a drifting Gabor presented to the left or right of the mouse subject. Stimulus difficulty is controlled by the contrast parameter, reported as a dimensionless Michelson fraction in $[0,1]$, where $1.0$ denotes full contrast and smaller values (e.g., $0.25$, $0.125$, $0.0625$) denote progressively harder stimuli.

To isolate how topology scales with stimulus strength, we group trials by experimental contrast levels $\{0.0625, 0.125, 0.25, 1.0\}$ and recompute the same metrics defined in Table~\ref{tab:mouse_metrics}: per-window mean AUBC ratio, Cohen’s $d$, and mutual information (MI). Results for low contrast ($0.0625$, $0.123$) can be seen in Table~\ref{tab:contrast_summary_low} and high contrast ($0.25$, $1.0$) can be seen in Table~\ref{tab:contrast_summary_high}.

\begin{table}[H]
  \centering
  \caption{Contrast-conditioned metrics (low contrasts). The mean AUBC ratio and Cohen's $d$ are shown for each window along with mutual information (MI).}
  \label{tab:contrast_summary_low}
  \small
  \setlength{\tabcolsep}{5pt}
  \renewcommand{\arraystretch}{1.15}
  \begin{tabular}{c|c|ccc|ccc}
    \toprule
    \multicolumn{2}{c|}{} &
    \multicolumn{3}{c|}{$\mathbf{0.0625}$ (n=34)} &
    \multicolumn{3}{c}{$\mathbf{0.125}$ (n=42)} \\
    \textbf{Dim} & \textbf{Window} &
    \textbf{Ratio} & \textbf{Cohen's $d$} & \textbf{MI} &
    \textbf{Ratio} & \textbf{Cohen's $d$} & \textbf{MI} \\
    \midrule
    \multirow{3}{*}{\textbf{0}} & Stimon
      & 1.3226 & 0.6953 & \multirow{3}{*}{0.4565}
      & 1.2416 & 0.6059 & \multirow{3}{*}{0.5833} \\
    & Trial
      & 1.1780 & 1.0718 &
      & 1.1896 & 0.8898 & \\
    & Feedback
      & 2.6557 & 2.0928 &
      & 2.7877 & 3.2266 & \\
    \midrule
    \multirow{3}{*}{\textbf{1}} & Stimon
      & 4.5424 & 1.8468 & \multirow{3}{*}{0.8213}
      & 5.5998 & 2.1874 & \multirow{3}{*}{0.9691} \\
    & Trial
      & 2.7822 & 1.4035 &
      & 4.3774 & 2.0010 & \\
    & Feedback
      & 17.7580 & 4.8979 &
      & 18.6113 & 6.5930 & \\
    \midrule
    \multirow{3}{*}{\textbf{2}} & Stimon
      & 2.5669 & 0.9967 & \multirow{3}{*}{0.5558}
      & 3.0085 & 1.1778 & \multirow{3}{*}{0.7772} \\
    & Trial
      & 1.8467 & 0.7923 &
      & 2.5555 & 1.4279 & \\
    & Feedback
      & 21.2726 & 1.6388 &
      & 24.7916 & 2.1093 & \\
    \midrule
    \multirow{3}{*}{\textbf{3}} & Stimon
      & 1.8789 & 0.7684 & \multirow{3}{*}{0.6696}
      & 2.4391 & 0.9573 & \multirow{3}{*}{0.6354} \\
    & Trial
      & 1.7179 & 1.1035 &
      & 2.2274 & 1.3431 & \\
    & Feedback
      & 11.6803 & 3.2099 &
      & 11.3447 & 3.5078 & \\
    \bottomrule
  \end{tabular}
\end{table}

\begin{table}[H]
  \centering
  \caption{Contrast-conditioned metrics (high contrasts). The mean AUBC ratio and Cohen's $d$ are shown for each window along with the mutual information (MI).}
  \label{tab:contrast_summary_high}
  \small
  \setlength{\tabcolsep}{5pt}
  \renewcommand{\arraystretch}{1.15}
  \begin{tabular}{c|c|ccc|ccc}
    \toprule
    \multicolumn{2}{c|}{} &
    \multicolumn{3}{c|}{$\mathbf{0.25}$ (n=38)} &
    \multicolumn{3}{c}{$\mathbf{1.0}$ (n=39)} \\
    \textbf{Dim} & \textbf{Window} &
    \textbf{Ratio} & \textbf{Cohen's $d$} & \textbf{MI} &
    \textbf{Ratio} & \textbf{Cohen's $d$} & \textbf{MI} \\
    \midrule
    \multirow{3}{*}{\textbf{0}} & Stimon
      & 1.2888 & 0.8082 & \multirow{3}{*}{0.5585}
      & 1.4801 & 0.9147 & \multirow{3}{*}{0.4894} \\
    & Trial
      & 1.3613 & 1.2612 &
      & 1.5481 & 1.4537 & \\
    & Feedback
      & 2.7507 & 2.6490 &
      & 2.7294 & 2.6329 & \\
    \midrule
    \multirow{3}{*}{\textbf{1}} & Stimon
      & 5.4215 & 2.2202 & \multirow{3}{*}{0.9537}
      & 6.6214 & 2.0069 & \multirow{3}{*}{1.0546} \\
    & Trial
      & 5.1011 & 2.5867 &
      & 6.8150 & 3.2881 & \\
    & Feedback
      & 17.0612 & 5.2312 &
      & 18.1621 & 7.2282 & \\
    \midrule
    \multirow{3}{*}{\textbf{2}} & Stimon
      & 2.8621 & 1.1903 & \multirow{3}{*}{0.6324}
      & 4.1155 & 1.3951 & \multirow{3}{*}{0.7883} \\
    & Trial
      & 2.8069 & 1.2768 &
      & 3.8631 & 1.3158 & \\
    & Feedback
      & 16.2966 & 1.7610 &
      & 18.3747 & 2.1431 & \\
    \midrule
    \multirow{3}{*}{\textbf{3}} & Stimon
      & 2.2864 & 0.9314 & \multirow{3}{*}{0.7241}
      & 2.9230 & 1.1129 & \multirow{3}{*}{0.7404} \\
    & Trial
      & 2.4194 & 2.1501 &
      & 2.9040 & 1.5450 & \\
    & Feedback
      & 10.4443 & 2.8083 &
      & 10.9396 & 4.5369 & \\
    \bottomrule
  \end{tabular}
\end{table}

Taken together, contrast modulates topological signal strength in a graded fashion. Across levels (0.0625, 0.125, 0.25, 1.0), AUBC ratios and Cohen’s $d$ increase with contrast and peak at Feedback, with dimension 1 showing the largest and most reliable effects (also the highest MI). Stimon and Trial scale similarly but remain smaller than Feedback. Higher Betti dimensions (2–3) exhibit very large ratios with more moderate Cohen's $d$/MI, consistent with low baselines.

\paragraph{Relation to aggregate activity.} Event-based increases in AUBC can naturally co-occur with elevated spiking. We therefore do not position this study as a superior technique for response-detection, the contribution is the structure of change: dimension-specific responses (e.g., a pronounced $d=1$ peak) and their graded modulation by contrast.

\section{Discussion}

Our results across synthetic and biological experiments underscore that emergent topological structure, captured through persistent homology on Transfer Entropy (TE) graphs, provides insightful representations of neural information processing. Importantly, these representations generalize robustly across tasks, complexity levels, and temporal regimes. The consistent emergence of task-specific topological fingerprints across diverse experimental settings suggests that neural systems dynamically reorganize their functional connectivity to meet computational demands.

In Experiment 1, focused on logic gate computations, we observed distinct topological profiles associated with each logic operation, particularly highlighting XOR's intrinsic complexity. While lower-dimensional topological features (dimensions 0–1) captured basic connectivity patterns similarly across gates, higher dimensions effectively discriminated XOR from simpler tasks (AND, OR). This aligns intuitively with XOR’s known non-linear separability, reflecting richer interactions and cyclic dependencies among neurons. The progressive dimensional differentiation further emphasizes persistent homology's sensitivity to subtle changes in computational complexity.

Experiment 2 expanded on this complexity theme by examining perceptual difficulty through structured perturbations of visual input. Both bit-flipping (noise addition) and bit-moving (pixel rearrangement) manipulations demonstrated that increasing perceptual challenge consistently elevates the complexity and variability of emergent network topology. Crucially, the preservation of total spike energy in the bit-moving scenario isolates topological changes as reflective of internal representational adjustments rather than mere input intensity changes. The smooth Wasserstein trajectories across noise levels further reveal that topological representations offer nuanced sensitivity to graded changes in stimulus complexity.

Experiment 3 bridged synthetic results to biological realism, illustrating the utility of this approach in capturing task-relevant neural dynamics from mouse cortical recordings. Behavioral epochs, particularly feedback phases, elicited pronounced increases in topological complexity, quantified robustly by metrics such as Cohen’s \textit{d}, relative baseline ratios, and mutual information. Notably, higher-dimensional features, despite their sparsity, were strongly informative of behavioral context, suggesting potential physiological relevance and functional specificity.

The varying sensitivities observed across homology dimensions offer additional insights into neural dynamics. Dimension 0 consistently provided fundamental connectivity information, dimension 1 robustly indexed feedback-driven changes, and higher dimensions (2–3) captured subtler but behaviorally relevant topological structures. Thus, examining multiple homology dimensions simultaneously enables a rich, hierarchical characterization of neural systems, ranging from broad network connectivity to intricate dynamical patterns, facilitating comprehensive interpretations of neural activity.

The separation of the trials into contrast classes demonstrated a clear relationship between contrast intensity and increased topological signatures relative to baseline. 

Moreover, the temporal resolution afforded by sliding window analysis in biological data demonstrate the method's suitability for tracking dynamic changes. The alignment of topological shifts with behavioral epochs strongly supports the interpretation of persistent homology as capturing  task related changes in neural interactions across various homology dimensions.

\section{Conclusion}

This paper presents a robust, interpretable, and generalizable pipeline for analyzing neural spiking dynamics through topological data analysis. By combining Transfer Entropy and directed persistent homology, we uncover distinct topological signatures reflective of underlying computational and behavioral contexts in both artificial and biological neural networks.

Across experiments involving logic gates, perturbed image classification, and mouse cortical activity, emergent topological structure consistently encoded complexity and task specificity. Notably, higher homology dimensions sensitively captured richer and subtler interaction patterns associated with computational difficulty. Our pipeline thus offers a methodological framework for deciphering the intricate dynamics of neural information flow. 

The demonstrated generalizability and interpretability suggest broad applicability in neuroscience and neural computation research, particularly in contexts demanding task-sensitive analyses of complex neural interactions. Future extensions could explore deeper integration with machine learning pipelines or application to larger-scale neural recordings, further bridging computational models and biological systems.

\subsection*{Acknowledgements} This research was supported by the Australian Government through the ARC's Discovery Projects funding scheme (project DP210103304). The first author was supported by a Research Training Program (RTP) Scholarship – Fee Offset by the Commonwealth Government.

\bibliographystyle{elsarticle-harv}
\bibliography{./literature.bib}

\begin{thebibliography}{37}
\expandafter\ifx\csname natexlab\endcsname\relax\def\natexlab#1{#1}\fi
\providecommand{\url}[1]{\texttt{#1}}
\providecommand{\href}[2]{#2}
\providecommand{\path}[1]{#1}
\providecommand{\DOIprefix}{doi:}
\providecommand{\ArXivprefix}{arXiv:}
\providecommand{\URLprefix}{URL: }
\providecommand{\Pubmedprefix}{pmid:}
\providecommand{\doi}[1]{\href{http://dx.doi.org/#1}{\path{#1}}}
\providecommand{\Pubmed}[1]{\href{pmid:#1}{\path{#1}}}
\providecommand{\bibinfo}[2]{#2}
\ifx\xfnm\relax \def\xfnm[#1]{\unskip,\space#1}\fi
%Type = Article
\bibitem[{Adams et~al.(2017)Adams, Emerson, Kirby, Neville, Peterson, Shipman,
  Chepushtanova, Hanson, Motta and Ziegelmeier}]{adams2017persistence}
\bibinfo{author}{Adams, H.}, \bibinfo{author}{Emerson, T.},
  \bibinfo{author}{Kirby, M.}, \bibinfo{author}{Neville, R.},
  \bibinfo{author}{Peterson, C.}, \bibinfo{author}{Shipman, P.},
  \bibinfo{author}{Chepushtanova, S.}, \bibinfo{author}{Hanson, E.},
  \bibinfo{author}{Motta, F.}, \bibinfo{author}{Ziegelmeier, L.},
  \bibinfo{year}{2017}.
\newblock \bibinfo{title}{Persistence images: A stable vector representation of
  persistent homology}.
\newblock \bibinfo{journal}{Journal of Machine Learning Research}
  \bibinfo{volume}{18}, \bibinfo{pages}{1--35}.
%Type = Inproceedings
\bibitem[{Bai et~al.(2024)Bai, Yu and Zhai}]{bai2024transitions}
\bibinfo{author}{Bai, X.}, \bibinfo{author}{Yu, C.}, \bibinfo{author}{Zhai,
  J.}, \bibinfo{year}{2024}.
\newblock \bibinfo{title}{Identifying the dynamical transitions of a stochastic
  spiking neural network by topological data analysis}, in:
  \bibinfo{booktitle}{Fourth International Conference on Advanced Algorithms
  and Neural Networks (AANN 2024)}, \bibinfo{organization}{SPIE}. pp.
  \bibinfo{pages}{360--364}.
%Type = Article
\bibitem[{Bardin et~al.(2019)Bardin, Spreemann and
  Hess}]{bardin2019topological}
\bibinfo{author}{Bardin, J.B.}, \bibinfo{author}{Spreemann, G.},
  \bibinfo{author}{Hess, K.}, \bibinfo{year}{2019}.
\newblock \bibinfo{title}{Topological exploration of artificial neuronal
  network dynamics}.
\newblock \bibinfo{journal}{Network Neuroscience} \bibinfo{volume}{3},
  \bibinfo{pages}{725--743}.
%Type = Article
\bibitem[{Beshkov et~al.(2024)Beshkov, Fyhn, Hafting and
  Einevoll}]{beshkov2024topological}
\bibinfo{author}{Beshkov, K.}, \bibinfo{author}{Fyhn, M.},
  \bibinfo{author}{Hafting, T.}, \bibinfo{author}{Einevoll, G.T.},
  \bibinfo{year}{2024}.
\newblock \bibinfo{title}{Topological structure of population activity in mouse
  visual cortex encodes densely sampled stimulus rotations}.
\newblock \bibinfo{journal}{Iscience} \bibinfo{volume}{27}.
%Type = Article
\bibitem[{Bubenik et~al.(2015)}]{bubenik2015statistical}
\bibinfo{author}{Bubenik, P.}, et~al., \bibinfo{year}{2015}.
\newblock \bibinfo{title}{Statistical topological data analysis using
  persistence landscapes.}
\newblock \bibinfo{journal}{J. Mach. Learn. Res.} \bibinfo{volume}{16},
  \bibinfo{pages}{77--102}.
%Type = Article
\bibitem[{Carlsson(2009)}]{carlsson2009topology}
\bibinfo{author}{Carlsson, G.}, \bibinfo{year}{2009}.
\newblock \bibinfo{title}{Topology and data}.
\newblock \bibinfo{journal}{Bulletin of the American Mathematical Society}
  \bibinfo{volume}{46}, \bibinfo{pages}{255--308}.
%Type = Inproceedings
\bibitem[{Chazal et~al.(2014)Chazal, Fasy, Lecci, Rinaldo and
  Wasserman}]{chazai2014sil}
\bibinfo{author}{Chazal, F.}, \bibinfo{author}{Fasy, B.T.},
  \bibinfo{author}{Lecci, F.}, \bibinfo{author}{Rinaldo, A.},
  \bibinfo{author}{Wasserman, L.}, \bibinfo{year}{2014}.
\newblock \bibinfo{title}{Stochastic convergence of persistence landscapes and
  silhouettes}, in: \bibinfo{booktitle}{Proceedings of the thirtieth annual
  symposium on Computational geometry}, \bibinfo{publisher}{Association for
  Computing Machinery}, \bibinfo{address}{New York, NY, USA}. p.
  \bibinfo{pages}{474–483}.
%Type = Article
\bibitem[{Chintakunta et~al.(2015)Chintakunta, Gentimis, Gonzalez-Diaz, Jimenez
  and Krim}]{CHINTAKUNTA2015391}
\bibinfo{author}{Chintakunta, H.}, \bibinfo{author}{Gentimis, T.},
  \bibinfo{author}{Gonzalez-Diaz, R.}, \bibinfo{author}{Jimenez, M.J.},
  \bibinfo{author}{Krim, H.}, \bibinfo{year}{2015}.
\newblock \bibinfo{title}{An entropy-based persistence barcode}.
\newblock \bibinfo{journal}{Pattern Recognition} \bibinfo{volume}{48},
  \bibinfo{pages}{391--401}.
%Type = Inproceedings
\bibitem[{Chowdhury and M\'{e}moli(2018)}]{chowdhury2018persistent}
\bibinfo{author}{Chowdhury, S.}, \bibinfo{author}{M\'{e}moli, F.},
  \bibinfo{year}{2018}.
\newblock \bibinfo{title}{Persistent path homology of directed networks}, in:
  \bibinfo{booktitle}{Proceedings of the Twenty-Ninth Annual ACM-SIAM Symposium
  on Discrete Algorithms}, \bibinfo{publisher}{Society for Industrial and
  Applied Mathematics}, \bibinfo{address}{USA}. p.
  \bibinfo{pages}{1152–1169}.
%Type = Article
\bibitem[{Cunningham and Yu(2014)}]{cunningham2014dimensionality}
\bibinfo{author}{Cunningham, J.P.}, \bibinfo{author}{Yu, B.M.},
  \bibinfo{year}{2014}.
\newblock \bibinfo{title}{Dimensionality reduction for large-scale neural
  recordings}.
\newblock \bibinfo{journal}{Nature neuroscience} \bibinfo{volume}{17},
  \bibinfo{pages}{1500--1509}.
%Type = Article
\bibitem[{Dabaghian et~al.(2012)Dabaghian, Mémoli, Frank and
  Carlsson}]{dabaghian2012topological}
\bibinfo{author}{Dabaghian, Y.}, \bibinfo{author}{Mémoli, F.},
  \bibinfo{author}{Frank, L.}, \bibinfo{author}{Carlsson, G.},
  \bibinfo{year}{2012}.
\newblock \bibinfo{title}{A topological paradigm for hippocampal spatial map
  formation using persistent homology}.
\newblock \bibinfo{journal}{PLOS Computational Biology} \bibinfo{volume}{8},
  \bibinfo{pages}{1--14}.
\newblock \DOIprefix\doi{10.1371/journal.pcbi.1002581}.
%Type = Book
\bibitem[{Edelsbrunner and Harer(2010)}]{EdelsbrunnerHarer2010}
\bibinfo{author}{Edelsbrunner, H.}, \bibinfo{author}{Harer, J.},
  \bibinfo{year}{2010}.
\newblock \bibinfo{title}{Computational Topology: An Introduction}.
\newblock Applied Mathematics, \bibinfo{publisher}{American Mathematical
  Society}.
%Type = Article
\bibitem[{Gardner et~al.(2022)Gardner, Hermansen, Pachitariu, Burak, Baas,
  Dunn, Moser and Moser}]{gardner2022toroidal}
\bibinfo{author}{Gardner, R.J.}, \bibinfo{author}{Hermansen, E.},
  \bibinfo{author}{Pachitariu, M.}, \bibinfo{author}{Burak, Y.},
  \bibinfo{author}{Baas, N.A.}, \bibinfo{author}{Dunn, B.A.},
  \bibinfo{author}{Moser, M.B.}, \bibinfo{author}{Moser, E.I.},
  \bibinfo{year}{2022}.
\newblock \bibinfo{title}{Toroidal topology of population activity in grid
  cells}.
\newblock \bibinfo{journal}{Nature} \bibinfo{volume}{602},
  \bibinfo{pages}{123--128}.
%Type = Book
\bibitem[{Gerstner and Kistler(2002)}]{gerstner2002spiking}
\bibinfo{author}{Gerstner, W.}, \bibinfo{author}{Kistler, W.M.},
  \bibinfo{year}{2002}.
\newblock \bibinfo{title}{Spiking neuron models: Single neurons, populations,
  plasticity}.
\newblock \bibinfo{publisher}{Cambridge university press}.
%Type = Article
\bibitem[{Ghosh-Dastidar and Adeli(2009)}]{ghosh2009spiking}
\bibinfo{author}{Ghosh-Dastidar, S.}, \bibinfo{author}{Adeli, H.},
  \bibinfo{year}{2009}.
\newblock \bibinfo{title}{Spiking neural networks}.
\newblock \bibinfo{journal}{International journal of neural systems}
  \bibinfo{volume}{19}, \bibinfo{pages}{295--308}.
%Type = Article
\bibitem[{Giusti et~al.(2015)Giusti, Pastalkova, Curto and
  Itskov}]{giusti2015clique}
\bibinfo{author}{Giusti, C.}, \bibinfo{author}{Pastalkova, E.},
  \bibinfo{author}{Curto, C.}, \bibinfo{author}{Itskov, V.},
  \bibinfo{year}{2015}.
\newblock \bibinfo{title}{Clique topology reveals intrinsic geometric structure
  in neural correlations}.
\newblock \bibinfo{journal}{Proceedings of the National Academy of Sciences}
  \bibinfo{volume}{112}, \bibinfo{pages}{13455--13460}.
%Type = Article
\bibitem[{Guidolin et~al.(2022)Guidolin, Desroches, Victor, Purpura and
  Rodrigues}]{guidolin2022geometry}
\bibinfo{author}{Guidolin, A.}, \bibinfo{author}{Desroches, M.},
  \bibinfo{author}{Victor, J.D.}, \bibinfo{author}{Purpura, K.P.},
  \bibinfo{author}{Rodrigues, S.}, \bibinfo{year}{2022}.
\newblock \bibinfo{title}{Geometry of spiking patterns in early visual cortex:
  a topological data analytic approach}.
\newblock \bibinfo{journal}{Journal of the Royal Society Interface}
  \bibinfo{volume}{19}, \bibinfo{pages}{20220677}.
%Type = Article
\bibitem[{Kang et~al.(2021)Kang, Xu and Morozov}]{kang2021cohomology}
\bibinfo{author}{Kang, L.}, \bibinfo{author}{Xu, B.}, \bibinfo{author}{Morozov,
  D.}, \bibinfo{year}{2021}.
\newblock \bibinfo{title}{Evaluating state space discovery by persistent
  cohomology in the spatial representation system}.
\newblock \bibinfo{journal}{Frontiers in computational neuroscience}
  \bibinfo{volume}{15}, \bibinfo{pages}{616748}.
%Type = Article
\bibitem[{L{\"u}tgehetmann et~al.(2020)L{\"u}tgehetmann, Govc, Smith and
  Levi}]{lutgehetmann2020computing}
\bibinfo{author}{L{\"u}tgehetmann, D.}, \bibinfo{author}{Govc, D.},
  \bibinfo{author}{Smith, J.P.}, \bibinfo{author}{Levi, R.},
  \bibinfo{year}{2020}.
\newblock \bibinfo{title}{Computing persistent homology of directed flag
  complexes}.
\newblock \bibinfo{journal}{Algorithms} \bibinfo{volume}{13},
  \bibinfo{pages}{19}.
%Type = Article
\bibitem[{Maass(1997)}]{maass1997networks}
\bibinfo{author}{Maass, W.}, \bibinfo{year}{1997}.
\newblock \bibinfo{title}{Networks of spiking neurons: the third generation of
  neural network models}.
\newblock \bibinfo{journal}{Neural networks} \bibinfo{volume}{10},
  \bibinfo{pages}{1659--1671}.
%Type = Article
\bibitem[{Moore et~al.(2018)Moore, Valentini, Walker and
  Levin}]{moore2018inform}
\bibinfo{author}{Moore, D.G.}, \bibinfo{author}{Valentini, G.},
  \bibinfo{author}{Walker, S.I.}, \bibinfo{author}{Levin, M.},
  \bibinfo{year}{2018}.
\newblock \bibinfo{title}{Inform: efficient information-theoretic analysis of
  collective behaviors}.
\newblock \bibinfo{journal}{Frontiers in Robotics and AI} \bibinfo{volume}{5},
  \bibinfo{pages}{60}.
%Type = Inproceedings
\bibitem[{Muller et~al.(2023)Muller, Kroon and Chalup}]{muller2023topological}
\bibinfo{author}{Muller, M.}, \bibinfo{author}{Kroon, S.},
  \bibinfo{author}{Chalup, S.}, \bibinfo{year}{2023}.
\newblock \bibinfo{title}{Topological dynamics of functional neural network
  graphs during reinforcement learning}, in: \bibinfo{booktitle}{International
  Conference on Neural Information Processing},
  \bibinfo{organization}{Springer}. pp. \bibinfo{pages}{190--204}.
%Type = Article
\bibitem[{Myers et~al.(2019)Myers, Munch and Khasawneh}]{myers2019persistent}
\bibinfo{author}{Myers, A.}, \bibinfo{author}{Munch, E.},
  \bibinfo{author}{Khasawneh, F.A.}, \bibinfo{year}{2019}.
\newblock \bibinfo{title}{Persistent homology of complex networks for dynamic
  state detection}.
\newblock \bibinfo{journal}{Physical Review E} \bibinfo{volume}{100},
  \bibinfo{pages}{022314}.
%Type = Article
\bibitem[{Naitzat et~al.(2020)Naitzat, Zhitnikov and Lim}]{naitzat2020topology}
\bibinfo{author}{Naitzat, G.}, \bibinfo{author}{Zhitnikov, A.},
  \bibinfo{author}{Lim, L.H.}, \bibinfo{year}{2020}.
\newblock \bibinfo{title}{Topology of deep neural networks}.
\newblock \bibinfo{journal}{Journal of Machine Learning Research}
  \bibinfo{volume}{21}, \bibinfo{pages}{1--40}.
%Type = Article
\bibitem[{Nardini et~al.(2021)Nardini, Stolz, Flores, Harrington and
  Byrne}]{nardini2021topological}
\bibinfo{author}{Nardini, J.T.}, \bibinfo{author}{Stolz, B.J.},
  \bibinfo{author}{Flores, K.B.}, \bibinfo{author}{Harrington, H.A.},
  \bibinfo{author}{Byrne, H.M.}, \bibinfo{year}{2021}.
\newblock \bibinfo{title}{Topological data analysis distinguishes parameter
  regimes in the anderson-chaplain model of angiogenesis}.
\newblock \bibinfo{journal}{PLOS Computational Biology} \bibinfo{volume}{17},
  \bibinfo{pages}{e1009094}.
%Type = Article
\bibitem[{Neftci et~al.(2019)Neftci, Mostafa and Zenke}]{neftci2019surrogate}
\bibinfo{author}{Neftci, E.O.}, \bibinfo{author}{Mostafa, H.},
  \bibinfo{author}{Zenke, F.}, \bibinfo{year}{2019}.
\newblock \bibinfo{title}{Surrogate gradient learning in spiking neural
  networks: Bringing the power of gradient-based optimization to spiking neural
  networks}.
\newblock \bibinfo{journal}{IEEE Signal Processing Magazine}
  \bibinfo{volume}{36}, \bibinfo{pages}{51--63}.
%Type = Article
\bibitem[{Otter et~al.(2017)Otter, Porter, Tillmann, Grindrod and
  Harrington}]{otter2017roadmap}
\bibinfo{author}{Otter, N.}, \bibinfo{author}{Porter, M.A.},
  \bibinfo{author}{Tillmann, U.}, \bibinfo{author}{Grindrod, P.},
  \bibinfo{author}{Harrington, H.A.}, \bibinfo{year}{2017}.
\newblock \bibinfo{title}{A roadmap for the computation of persistent
  homology}.
\newblock \bibinfo{journal}{EPJ Data Science} \bibinfo{volume}{6},
  \bibinfo{pages}{1--38}.
%Type = Article
\bibitem[{Perea and Harer(2015)}]{perea2015sliding}
\bibinfo{author}{Perea, J.A.}, \bibinfo{author}{Harer, J.},
  \bibinfo{year}{2015}.
\newblock \bibinfo{title}{Sliding windows and persistence: An application of
  topological methods to signal analysis}.
\newblock \bibinfo{journal}{Foundations of computational mathematics}
  \bibinfo{volume}{15}, \bibinfo{pages}{799--838}.
%Type = Article
\bibitem[{Petri et~al.(2014)Petri, Expert, Turkheimer, Carhart-Harris, Nutt,
  Hellyer and Vaccarino}]{petri2014homological}
\bibinfo{author}{Petri, G.}, \bibinfo{author}{Expert, P.},
  \bibinfo{author}{Turkheimer, F.}, \bibinfo{author}{Carhart-Harris, R.},
  \bibinfo{author}{Nutt, D.}, \bibinfo{author}{Hellyer, P.J.},
  \bibinfo{author}{Vaccarino, F.}, \bibinfo{year}{2014}.
\newblock \bibinfo{title}{Homological scaffolds of brain functional networks}.
\newblock \bibinfo{journal}{Journal of The Royal Society Interface}
  \bibinfo{volume}{11}, \bibinfo{pages}{20140873}.
%Type = Article
\bibitem[{Reimann et~al.(2017)Reimann, Nolte, Scolamiero, Turner, Perin,
  Chindemi, D{\l}otko, Levi, Hess and Markram}]{reimann2017cliques}
\bibinfo{author}{Reimann, M.W.}, \bibinfo{author}{Nolte, M.},
  \bibinfo{author}{Scolamiero, M.}, \bibinfo{author}{Turner, K.},
  \bibinfo{author}{Perin, R.}, \bibinfo{author}{Chindemi, G.},
  \bibinfo{author}{D{\l}otko, P.}, \bibinfo{author}{Levi, R.},
  \bibinfo{author}{Hess, K.}, \bibinfo{author}{Markram, H.},
  \bibinfo{year}{2017}.
\newblock \bibinfo{title}{Cliques of neurons bound into cavities provide a
  missing link between structure and function}.
\newblock \bibinfo{journal}{Frontiers in computational neuroscience}
  \bibinfo{volume}{11}, \bibinfo{pages}{266051}.
%Type = Article
\bibitem[{Saggar et~al.(2018)Saggar, Sporns, Gonzalez-Castillo, Bandettini,
  Carlsson, Glover and Reiss}]{saggar2018dynamical}
\bibinfo{author}{Saggar, M.}, \bibinfo{author}{Sporns, O.},
  \bibinfo{author}{Gonzalez-Castillo, J.}, \bibinfo{author}{Bandettini, P.A.},
  \bibinfo{author}{Carlsson, G.}, \bibinfo{author}{Glover, G.},
  \bibinfo{author}{Reiss, A.L.}, \bibinfo{year}{2018}.
\newblock \bibinfo{title}{Towards a new approach to reveal dynamical
  organization of the brain using topological data analysis}.
\newblock \bibinfo{journal}{Nature communications} \bibinfo{volume}{9},
  \bibinfo{pages}{1399}.
%Type = Article
\bibitem[{Santos et~al.(2019)Santos, Raposo, Coutinho-Filho, Copelli, Stam and
  Douw}]{santos2019topological}
\bibinfo{author}{Santos, F.A.}, \bibinfo{author}{Raposo, E.P.},
  \bibinfo{author}{Coutinho-Filho, M.D.}, \bibinfo{author}{Copelli, M.},
  \bibinfo{author}{Stam, C.J.}, \bibinfo{author}{Douw, L.},
  \bibinfo{year}{2019}.
\newblock \bibinfo{title}{Topological phase transitions in functional brain
  networks}.
\newblock \bibinfo{journal}{Physical Review E} \bibinfo{volume}{100},
  \bibinfo{pages}{032414}.
%Type = Article
\bibitem[{Schreiber(2000)}]{schreiber2000measuring}
\bibinfo{author}{Schreiber, T.}, \bibinfo{year}{2000}.
\newblock \bibinfo{title}{Measuring information transfer}.
\newblock \bibinfo{journal}{Physical review letters} \bibinfo{volume}{85},
  \bibinfo{pages}{461}.
%Type = Article
\bibitem[{Sizemore et~al.(2019)Sizemore, Phillips-Cremins, Ghrist and
  Bassett}]{sizemore2019importance}
\bibinfo{author}{Sizemore, A.E.}, \bibinfo{author}{Phillips-Cremins, J.E.},
  \bibinfo{author}{Ghrist, R.}, \bibinfo{author}{Bassett, D.S.},
  \bibinfo{year}{2019}.
\newblock \bibinfo{title}{The importance of the whole: topological data
  analysis for the network neuroscientist}.
\newblock \bibinfo{journal}{Network Neuroscience} \bibinfo{volume}{3},
  \bibinfo{pages}{656--673}.
%Type = Article
\bibitem[{Stolz et~al.(2017)Stolz, Harrington and Porter}]{stolz2017persistent}
\bibinfo{author}{Stolz, B.J.}, \bibinfo{author}{Harrington, H.A.},
  \bibinfo{author}{Porter, M.A.}, \bibinfo{year}{2017}.
\newblock \bibinfo{title}{Persistent homology of time-dependent functional
  networks constructed from coupled time series}.
\newblock \bibinfo{journal}{Chaos: An Interdisciplinary Journal of Nonlinear
  Science} \bibinfo{volume}{27}, \bibinfo{pages}{047410}.
%Type = Article
\bibitem[{{The International Brain Laboratory} et~al.(2021){The International
  Brain Laboratory}, Aguillon-Rodriguez, Angelaki, Bayer, Bonacchi, Carandini,
  Cazettes, Chapuis, Churchland, Dan, Dewitt, Faulkner, Forrest, Haetzel,
  Häusser, Hofer, Hu, Khanal, Krasniak, Laranjeira, Mainen, Meijer, Miska,
  Mrsic-Flogel, Murakami, Noel, Pan-Vazquez, Rossant, Sanders, Socha, Terry,
  Urai, Vergara, Wells, Wilson, Witten, Wool and Zador}]{ibl2021neuro}
\bibinfo{author}{{The International Brain Laboratory}},
  \bibinfo{author}{Aguillon-Rodriguez, V.}, \bibinfo{author}{Angelaki, D.},
  \bibinfo{author}{Bayer, H.}, \bibinfo{author}{Bonacchi, N.},
  \bibinfo{author}{Carandini, M.}, \bibinfo{author}{Cazettes, F.},
  \bibinfo{author}{Chapuis, G.}, \bibinfo{author}{Churchland, A.K.},
  \bibinfo{author}{Dan, Y.}, \bibinfo{author}{Dewitt, E.},
  \bibinfo{author}{Faulkner, M.}, \bibinfo{author}{Forrest, H.},
  \bibinfo{author}{Haetzel, L.}, \bibinfo{author}{Häusser, M.},
  \bibinfo{author}{Hofer, S.B.}, \bibinfo{author}{Hu, F.},
  \bibinfo{author}{Khanal, A.}, \bibinfo{author}{Krasniak, C.},
  \bibinfo{author}{Laranjeira, I.}, \bibinfo{author}{Mainen, Z.F.},
  \bibinfo{author}{Meijer, G.}, \bibinfo{author}{Miska, N.J.},
  \bibinfo{author}{Mrsic-Flogel, T.D.}, \bibinfo{author}{Murakami, M.},
  \bibinfo{author}{Noel, J.P.}, \bibinfo{author}{Pan-Vazquez, A.},
  \bibinfo{author}{Rossant, C.}, \bibinfo{author}{Sanders, J.},
  \bibinfo{author}{Socha, K.}, \bibinfo{author}{Terry, R.},
  \bibinfo{author}{Urai, A.E.}, \bibinfo{author}{Vergara, H.},
  \bibinfo{author}{Wells, M.}, \bibinfo{author}{Wilson, C.J.},
  \bibinfo{author}{Witten, I.B.}, \bibinfo{author}{Wool, L.E.},
  \bibinfo{author}{Zador, A.M.}, \bibinfo{year}{2021}.
\newblock \bibinfo{title}{Standardized and reproducible measurement of
  decision-making in mice}.
\newblock \bibinfo{journal}{eLife} \bibinfo{volume}{10},
  \bibinfo{pages}{e63711}.
%Type = Article
\bibitem[{Xi et~al.(2025)Xi, Fan, Wang, Li and Yang}]{XI2025130086}
\bibinfo{author}{Xi, X.}, \bibinfo{author}{Fan, Z.}, \bibinfo{author}{Wang,
  T.}, \bibinfo{author}{Li, L.}, \bibinfo{author}{Yang, J.},
  \bibinfo{year}{2025}.
\newblock \bibinfo{title}{Topology analysis of eeg-based functional brain
  network after stroke}.
\newblock \bibinfo{journal}{Neurocomputing} \bibinfo{volume}{637},
  \bibinfo{pages}{130086}.

\end{thebibliography}

\end{document}